\documentclass{aa}

\usepackage{graphicx}
\usepackage{txfonts}
\usepackage{natbib}
\bibpunct{(}{)}{;}{a}{}{,}

\def\harrow{\mathrel{\hbox{\rlap{\hbox{\raise4pt\hbox{${\rm +H}$}}}\hbox{$\longrightarrow$}}}}
\def\oarrow{\mathrel{\hbox{\rlap{\hbox{\raise4pt\hbox{${\rm +O}$}}}\hbox{$\longrightarrow$}}}}

\def\lesssim{\mathrel{\hbox{\rlap{\hbox{\lower4pt\hbox{$\sim$}}}\hbox{$<$}}}}
\def\gtsim{\mathrel{\hbox{\rlap{\hbox{\lower4pt\hbox{$\sim$}}}\hbox{$>$}}}}

\def\farcs{\hbox{$.\!\!^{\prime\prime}$}}

\newcommand{\kms}          	{\mbox{${\rm km~s^{-1}}$}}
\newcommand{\ccs}         		{\mbox{${\rm cm^3~s^{-1}}$}}
\newcommand{\cc}         		{\mbox{${\rm cm^{-3}}$}}
\newcommand{\hdensity}         	{n_{{\rm H}_2}}

\newcommand{\fdh}        		{\mbox{${\rm H_2CO}$}}
\newcommand{\m}          		{\mbox{${\rm CH_3OH}$}}
\newcommand{\ammonia}         {\mbox{${\rm NH_3}$}}
\newcommand{\mf}          		{\mbox{${\rm HCOOCH_3}$}}
\newcommand{\dime}          	{\mbox{${\rm CH_3OCH_3}$}}
\newcommand{\mc}         		{\mbox{${\rm CH_3CN}$}}
\newcommand{\ec}        		{\mbox{${\rm C_2H_5CN}$}}

\begin{document}

\title{Hot corinos in NGC1333-IRAS4B and IRAS2A}
\author{S. Bottinelli\inst{1,2} \and C. Ceccarelli\inst{1} 
\and J. P. Williams\inst{2} \and B. Lefloch\inst{1}
}

\offprints{S. Bottinelli}

\institute{
Laboratoire d'Astrophysique de l'Observatoire de Grenoble, 
BP 53, 38041 Grenoble, Cedex 9, France. \\
\email{sbottine,ceccarel,lefloch@obs.ujf-grenoble.fr}
\and Institute for Astronomy, University of Hawai`i, 
2680 Woodlawn Drive, Honolulu HI 96822, USA.\\
\email{jpw@ifa.hawaii.edu}
}

\date{Received ; Accepted October, 11 2006}

\abstract{Complex organic molecules have been detected in massive hot
   cores for over two decades, and only recently in three hot corinos
   (the inner regions surrounding Sun-like protostars,
   where the dust temperature exceeds 100\,K).
   Since hot corinos have sizes of $\sim$100\,AU (i.e., of the order of the extent
   of the Solar System), it is particularly relevant to understand
   whether they are common and to identify the formation route(s) of
   complex organic molecules. Much has yet to be learned on this
   topic, since even recent models predicted it was not possible to
   form these molecules in low-mass protostars.}
   {We aim to enlarge the number of known hot corinos and carry out a
   first comparative study with hot cores.  The ultimate goal is to
   understand whether complex organic molecules form in the gas phase
   or on grain surfaces, and what the possible key parameters are.}
   {We observed millimeter rotational transitions of HCOOH, \mf\,
   \dime\, \mc\, and \ec\ in a sample of low-mass protostars with the
   IRAM-30m. Using the rotational diagram method coupled with the
   information about the sources' structure, we calculate the
   abundances of the observed molecules. To interpret these
   abundances, we review the proposed formation processes of the above
   molecules.}
   {We report the detection of \mf\ and/or \mc\ towards NGC1333-IRAS4B
   and NGC1333-IRAS2A.  We find that abundance ratios of O-bearing
   molecules to methanol or formaldehyde in hot corinos are comparable
   and about unity, and are relatively (depending on how the
   ratios are determined) higher than those in hot cores and in
   Galactic center clouds.}
   {So far, complex organic molecules were detected in all the hot
   corinos where they were searched for, suggesting that it is a
   common phase for low-mass protostars. While some evidence points to
   grain-surface synthesis (either in the cold or warm-up phase)
   of these molecules (in particular for HCOOH and \mf),
   the present data do not allow us to
   disregard gas-phase formation. More observational, laboratory, and
   theoretical studies are required to improve our understanding of
   hot corinos.}


%\titlerunning{}
%\authorrunning{S. Bottinelli et al.}

\maketitle

\section{Introduction}

Many aspects of the formation of solar-type stars have now been
elucidated and, even though the details are much debated, there exists
a widely accepted framework for it (see the volume Protostars and
Planets V, for example, 
\citealt{ceccarelli-etal06,ward-thompson-etal06-ppv,white-etal06}). 
Low-mass stars form inside molecular clouds
from dense and cold condensations, called prestellar cores, which
evolve into Class 0 and then Class I sources. In the latest phases,
the newly born star is surrounded by a proto-planetary disk, which
eventually may form planets.

Somewhat less understood, and therefore more frequently debated, is the chemical
evolution of matter from the molecular cloud to the proto-planetary
phase, and then to planets (e.g., Ceccarelli et al. 2006).  It is now
acknowledged that, at the densities and temperatures typical of the
centers of prestellar cores ($\geq 10^6$\,cm$^{-3}$ and $\leq 10$\,K,
respectively), heavy-element bearing molecules condense out onto dust
grains, forming icy mantles. Very likely, hydrogenation and oxidation
of small molecules and atoms (like CO, O, N, etc.) occur on the grain
surfaces, so that the mantles end up being composed mainly of water,
interspersed with traces of formaldehyde, methanol, ammonia, and
possibly even more complex molecules.  As material from the
surrounding envelope starts accreting onto the central protostar, the
increased radiation output heats up the surroundings.
When and where the dust reaches the appropriate sublimation
temperature of the ices (which depends on the exact composition of
these ices), the grain mantles evaporate, injecting their components
into the gas phase. These components can further react to form
more complex molecules. The regions where the ice mantles sublimate (at
T$_{\rm dust}\gtsim$ 100 K) and where the emission from complex
molecules (whether evaporated from the grain mantles or formed in the gas)
originates, are called hot corinos
\citep{ceccarelli04,bottinelli-etal04-iras4a}. The chemical
composition of these hot corinos reflects both the heritage from the
prestellar core phase and the reactions taking place in the warm
gas. The result is a gas rich in complex organic molecules
\citep{cazaux-etal03}, whose diversity is far from being fully
explored and understood. It is likely that we have discovered only the tip
of the iceberg. Moreover, the story does not end here. Indeed, the
fate of these complex organic molecules is almost totally
unknown. They may condense onto the grain surfaces again during
the proto-planetary disk phase.  Perhaps they are incorporated into
the planetesimals forming the building blocks of planets, or into
comets and asteroids, in which case they may end up on newly formed
planets as accretion proceeds.
This picture is, at least partially, supported by the fact that some
complex organic molecules are found both in hot corinos and in comets
(e.g., methyl cyanide \mc, methyl formate \mf, and formic acid HCOOH;
\citealt{ehrenfreund+charnley00}). For these reasons, the hot corino
phase is not only interesting in itself, but it is a critical phase in
the process of solar-type star formation.

In short, the molecular complexity in hot corinos is particularly
relevant to molecular astrophysics and (exo)planetary science, but
particularly unknown. Our ignorance can be summarized by this simple,
multiple-part question: {\it  which complex organic molecules are formed, 
and why, where, and how does this take place?}
At present we are simply unable to answer this
question for the following reasons.  {\it (i)} We do not have a census
of the complex organic molecules. We suspect that many complex
molecules are present in the millimeter spectra of hot corinos, but
the firm identification of a large molecule requires several lines
from this molecule and can be hampered by the multitude of weak lines
in the spectra
\citep[e.g.,][]{combes-etal96,ceccarelli-etal00-h2co16293}.  {\it (ii)}
We do not know why they form. The gas-grain models for hot cores (the
high-mass analogs of hot corinos) predict that there is not enough
time for hot corinos to form molecules because the gas falls rapidly
towards the central forming star before any gas-phase reaction can
lead to complex organic molecules \citep[e.g.,][]{schoier-etal02}. {\it
(iii)} We debate whether the observed complex molecules reside in the
passive heated envelope \citep{ceccarelli-etal00-structure16293,
ceccarelli-etal00-glycine16293}, in the hidden circumstellar disk
\citep{jorgensen-etal05-h2co-ch3oh}, or in the interface between the
outflow and the envelope \citep{chandler-etal05}. {\it (iv)} Finally,
we do not know how complex molecules form: in the gas-phase or on the
grain surfaces, or both. Much of our ignorance stems from the too few
hot corinos so far studied: IRAS16293--2422 (hereafter IRAS16293, 
\citealt{cazaux-etal03,kuan-etal04,bottinelli-etal04-iras16293,remijan+hollis06}),
NGC1333-IRAS4A (hereafter IRAS4A, \citealt{bottinelli-etal04-iras4a}),
NGC1333-IRAS4B (hereafter IRAS4B, \citealt{sakai-etal06}),
and NGC1333-IRAS2A (hereafter
IRAS2A, \citealt{jorgensen-etal05-iras2a}).\\

In this paper, we present detections of complex organic molecules in
two Class 0 hot corinos: \mf\ and \mc\ in 
IRAS4B, and \mc\ in IRAS2A. The article is organized as
follows: details on the choice of sources are given in Sect.
\ref{background}, the observations are described in Sect.
\ref{observations} and results are presented and discussed in Sects.
\ref{results} and \ref{discussion}, respectively.  Section
\ref{discussion} starts with 
some remarks on the formation routes of complex organic
molecules (Sect. \ref{summary-formation}, based on
a summary of the possible formation mechanisms of
the complex molecules HCOOH, \mf, \dime, \mc, and \ec\
given in Appendix \ref{formation}), followed by the analysis of our data
(Sects. \ref{luminosity} and \ref{abundance-ratios}), and ends with a
comparison of hot corinos with hot cores (Sect. \ref{comparison}) and with
Galactic center clouds (Sect. \ref{gc}).  Finally, a summary and concluding
remarks are given in Sect. \ref{conclusion}.

\section{Source selection and background \label{background}}

We observed two Class 0 protostars, IRAS4B and IRAS2A, located in the
Perseus complex, specifically in the NGC1333 cloud, whose distance is
estimated to be 220 pc \citep{cernis90}.  IRAS4B belongs to the
multiple system IRAS4 and is located $\sim 30''$ and $\sim 17''$ from
the other two components, IRAS4A and IRAS4C
\citep{looney-etal00,reipurth-etal02}. It may be a multiple
stellar system itself \citep{looney-etal00}. Both IRAS4A and IRAS4B are
associated with molecular outflows of a dynamical age of a few
thousand years, seen in CO, CS \citep{blake-etal95}, and SiO
\citep{lefloch-etal98}.  IRAS2A is part of a protobinary system and is
separated by 30$''$ from its companion IRAS2B
\citep{looney-etal00,reipurth-etal02}.  Two CO bipolar outflows appear
to originate within a few arcsec of IRAS2A: a highly collimated jet in
the east-west direction and a large-scale outflow aligned NNE-SSW
\citep{sandell-etal94,knee+sandell00}.\\

IRAS4B and IRAS2A were selected from the sample studied in
\citet{maret-etal04,maret-etal05} because they both are good hot
corino candidates and because, due to their distance and luminosity,
they are expected to have brighter lines compared to other Class 0
sources (e.g., from the \citealt{andre-etal00} sample). The candidacy
relies upon the claim by \citet{maret-etal04,maret-etal05} of the
presence, in both IRAS2A and IRAS4B, of a warm ($\gtsim 100$ K) inner
region where grain mantles sublimate.  The claim is based on the
observed jumps in the abundances of formaldehyde and methanol: low
abundances in the outer, cold envelope where formaldehyde and methanol
are still frozen onto grain surfaces, and high abundances in the
inner, warm envelope where the heat from the central object causes the
desorption of formaldehyde and methanol into the gas phase.
Indeed, in the two sources where hot corinos have been 
detected so far, IRAS16293 \citep{cazaux-etal03} and IRAS4A
\citep{bottinelli-etal04-iras4a}, similar jumps in the formaldehyde
and/or methanol abundances have also been claimed
\citep{maret-etal04,maret-etal05}, supporting the choice of our
targets. \\

However, there is a noticeable difference in the extent of
the hot corinos (54 to 266 AU, \citealt{maret-etal04}) 
and in the jump sizes in all four sources.
Indeed, according to \citet{maret-etal05}, methanol abundances show
jumps of a factor 100 to 350 in all sources but IRAS4A, where the jump
is lower than about a factor 15, whereas \citet{maret-etal04} found
formaldehyde abundance jumps of a factor 100 to 6000.  Other authors
have carried out analyses of methanol and formaldehyde in low-mass
protostars \citep[e.g.,][]{schoier-etal02,jorgensen-etal05-h2co-ch3oh}.
In contrast with what was found by \citet{maret-etal04},
\citet{jorgensen-etal05-h2co-ch3oh} claimed that no jump of
formaldehyde abundance is required to model the line intensities in
any of the sources except IRAS16293 \citep{schoier-etal02}.  However,
like \citet{maret-etal05}, \citet{jorgensen-etal05-h2co-ch3oh} 
and \citet{schoier-etal02} 
found a methanol abundance jump in IRAS4B,
IRAS2A, and IRAS16293, but not necessarily in IRAS4A.  The hot corino
abundances of formaldehyde and methanol found by the two groups are
summarized in Table \ref{tab:sources}.  In the following we will adopt
the \citet{maret-etal04,maret-etal05} framework, but we will also
discuss the results in the light of the
\citet{jorgensen-etal05-h2co-ch3oh} analysis.\\

In summary, the four hot corinos --- IRAS16293, IRAS4A, IRAS4B, and
IRAS2A --- form a sample of interestingly different sources, both
because they span an apparent large range of formaldehyde and methanol
abundances, and because the very existence of the abundance jumps is
still debated.
Hence, even though statistically small, this sample will permit a
first assessment of how the presence and abundance of complex organic
molecules in hot corinos depend on formaldehyde and methanol.  Indeed,
these two molecules are predicted to be among the most important
parent ones \citep[e.g.,][]{caselli-etal93,
rodgers+charnley01,rodgers+charnley03}, but it is not always clear how
they relate to the complex molecules \citep[e.g.,][]{horn-etal04}.

\begin{table*}
\caption{Hot corino abundances for formaldehyde ($X_{\rm hc}$(H$_2$CO)) and methanol ($X_{\rm hc}$(CH$_3$OH)).}
\label{tab:sources}
\begin{scriptsize}
\begin{center}
 \begin{tabular}{lrccccc}
 \hline\hline
Source & $L_{\rm bol}$$^a$ & \multicolumn{2}{c}{$X_{\rm hc}$(H$_2$CO)}   & & \multicolumn{2}{c}{$X_{\rm hc}$(CH$_3$OH)} \\
\cline{3-4}\cline{6-7} 
        &     ($L_\odot$)   &  \citet{maret-etal04} & \citet{jorgensen-etal05-h2co-ch3oh}$^b$ & & \citet{maret-etal05} &  \citet{jorgensen-etal05-h2co-ch3oh}$^c$ \\
 \hline
 
 IRAS16293--2422& 27 & 1$\times10^{-7}$ & 6$\times10^{-8}$$^d$  & &  1$\times10^{-7}$ & 1$\times10^{-7}$\\ 
 NGC1333-IRAS4A & 6 & 2$\times10^{-8}$ & $3\times10^{-9}$ & & $<1\times10^{-8}$ & $\leq3.5\times10^{-9}$\\
 NGC1333-IRAS4B & 6 & 3$\times10^{-6}$ & $1\times10^{-8}$ & &  7$\times10^{-7}$ & 9.5$\times10^{-8}$\\
 NGC1333-IRAS2A & 16 & 2$\times10^{-7}$ &  $8\times10^{-10}$ & &  3$\times10^{-7}$ & 1.5$\times10^{-7}$\\
 \hline
 \end{tabular}
\end{center}

$^a$ From \citet{jorgensen-etal02}.\\
$^b$ Note that except in IRAS16293 \citep{schoier-etal02}, the formaldehyde abundances modeled by \citet{jorgensen-etal05-h2co-ch3oh} do not require any jump (see text for further details).\\
$^c$ Methanol abundances averaged over A- and E-types.\\
$^d$ From \citet{schoier-etal02}.\\
\end{scriptsize}
\end{table*}

\section{Observations \label{observations}}

The observations were carried out in June 2003 with the 30-meter telescope
of the Institut de RadioAstronomie Millim\'etrique (IRAM)\footnote{
IRAM is an international venture supported by INSU/CNRS (France), MPG (Germany),
and IGN (Spain).}.
The positions used for pointing were
($\alpha$, $\delta$)(2000) = ($03^{\rm h}29^{\rm m}12\fs 0$, 
$31^\circ 13'09''$) for IRAS4B, and
($\alpha$, $\delta$)(2000) = ($03^{\rm h}28^{\rm m}55\fs 4$,
$31^\circ 14'35''$) for IRAS2A.
Based on the observations of IRAS16293 by \citet{cazaux-etal03}, we
targeted the following complex molecules: methyl formate, HCOOCH$_3$
(A and E), formic acid, HCOOH, dimethyl ether, CH$_3$OCH$_3$, methyl
cyanide, CH$_3$CN, and ethyl cyanide, C$_2$H$_5$CN. Different
telescope settings were used to include as many transitions
as possible for each molecule (see Table \ref{setups}). All lines were
observed with a low resolution, 1~MHz filter bank of 4~$\times$~256
channels split between different receivers, providing a velocity
resolution of $\sim$ 3, 2, and 1~\kms\ at 3, 2, and 1~mm, respectively. 
Each receiver was simultaneously connected to a unit of
the autocorrelator, with spectral resolutions of 20, 80, or 320 kHz and
bandwidths between 40 and 240 MHz, equivalent to an (unsmoothed)
velocity resolution of 0.1--0.4~\kms\ at 3, 2, and 1~mm.
Typical system temperatures were 100--200~K,
180--250~K, and 500--1500~K, at 3, 2, and 1~mm, respectively.

\begin{table}
\caption{Observed frequencies and targeted molecules.}
\label{setups}
\begin{center}
\begin{scriptsize}
\begin{tabular}{cllcc}
\hline\hline
Frequency range & Molecule & Frequencies  & rms (mK)  & rms (mK) \\
         (GHz)  &          &   (GHz)      & IRAS4B    &  IRAS2A  \\
\hline
 90.07 --  90.32    & HCOOCH$_3$-A  & 90.156, 90.229            & 7     & 2\\
                               & HCOOCH$_3$-E  & 90.145, 90.227            & 2, 7 & 2\\
                               & HCOOH                   & 90.164                          & 2     & 2\\
 98.50 --  98.75    & HCOOCH$_3$-A  & 98.611, 98.682            & 7, 2 & 3\\
                               & HCOOCH$_3$-E  & 98.606, 98.711            & 7, 2 & 3\\
                               & C$_2$H$_5$CN   & 98.610, 98.701            & 7, 2 & 3\\
110.29 -- 110.43 & CH$_3$CN             & 110.329 -- 110.383    & 5     & -- \\
223.17 -- 223.42 & CH$_3$OCH$_3$ & 223.200 -- 223.204    & 13   & 34\\
                               & C$_2$H$_5$CN   & 223.385                         & 13   & 34\\
226.50 -- 227.00 & HCOOCH$_3$-A   & 226.718, 226.778        & 11   & 15\\
                               & HCOOCH$_3$-E   & 226.713, 226.773        & 11   & 15\\
257.35 -- 257.55 & CH$_3$CN             & 257.403 -- 257.527      & 22   & 28\\
\hline
\end{tabular}
\end{scriptsize}
\end{center}
{\sc Note}. -- A dash indicates that no data were taken in the corresponding
frequency range.
\end{table}

\begin{figure*}
\includegraphics[width=0.5\textwidth]{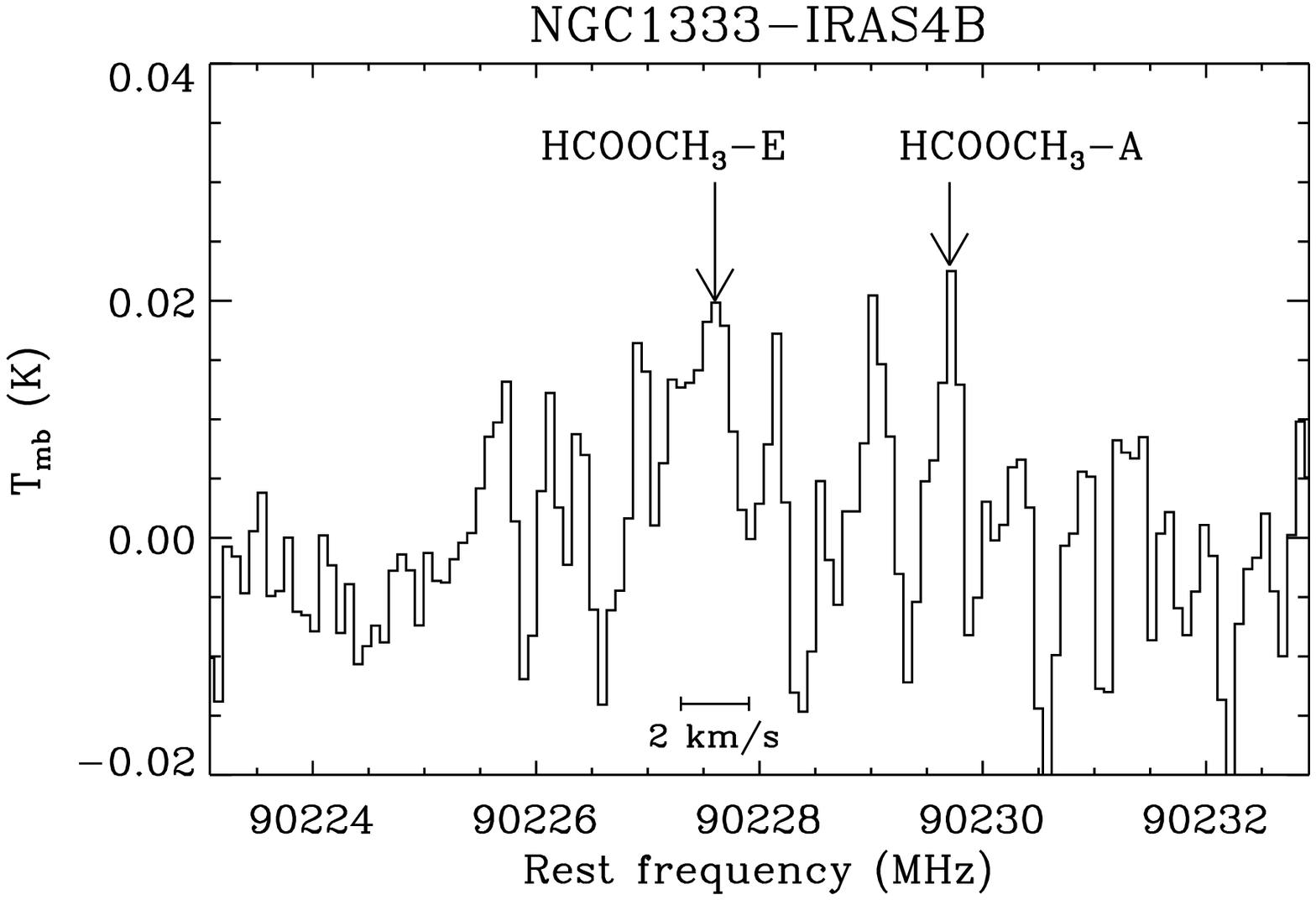}
\includegraphics[width=0.5\textwidth]{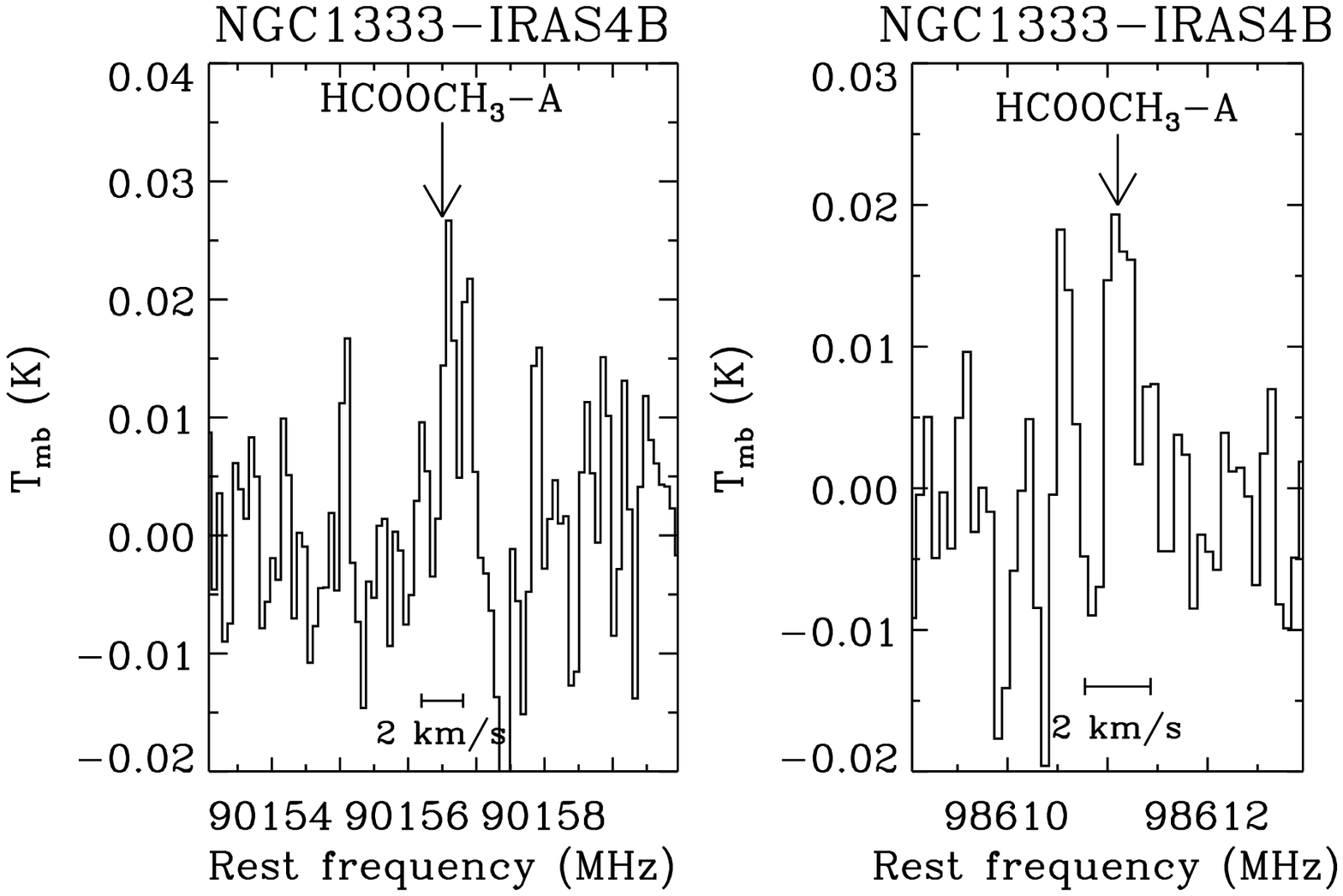}
\includegraphics[width=0.5\textwidth]{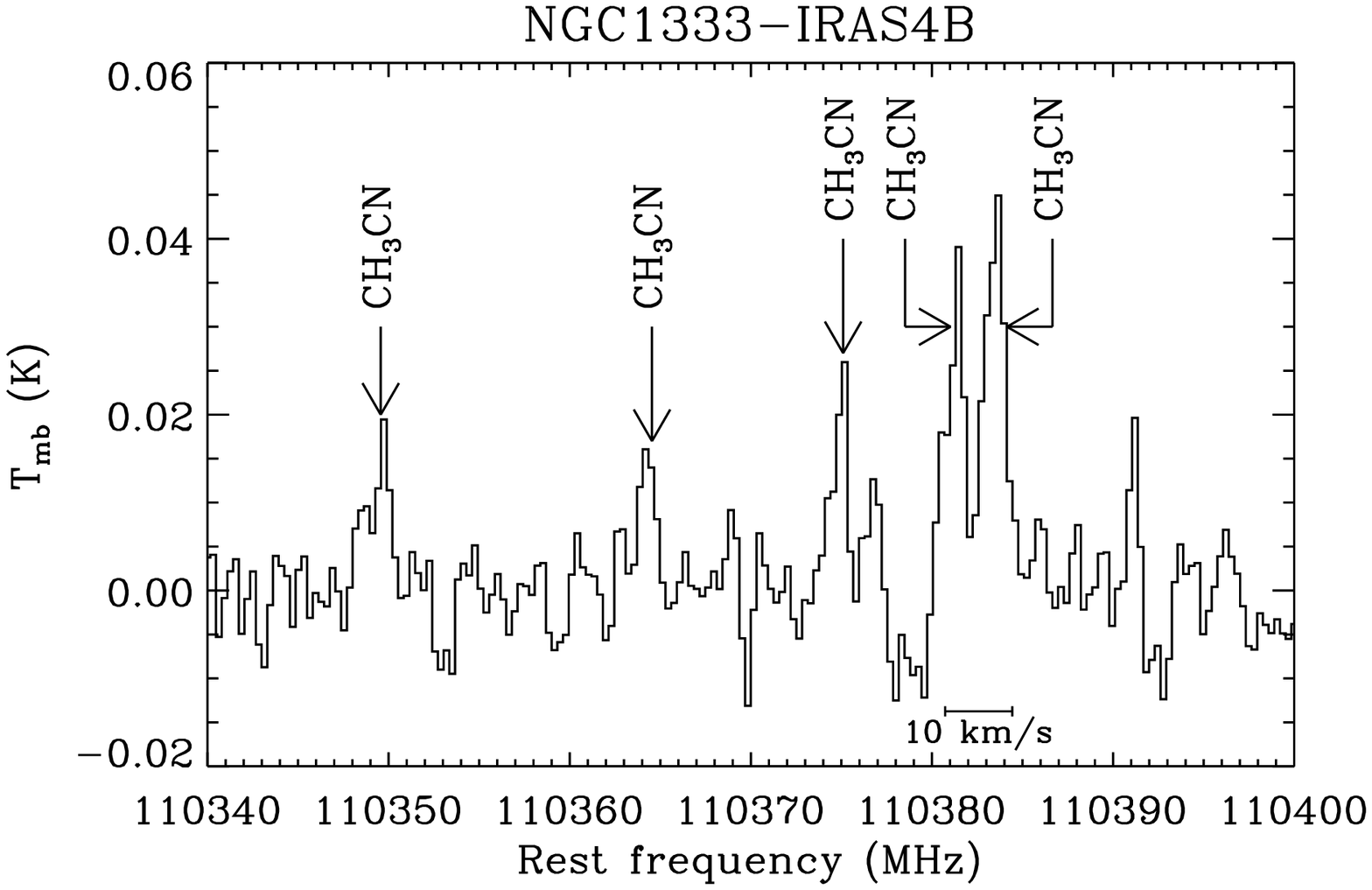}
\includegraphics[width=0.5\textwidth]{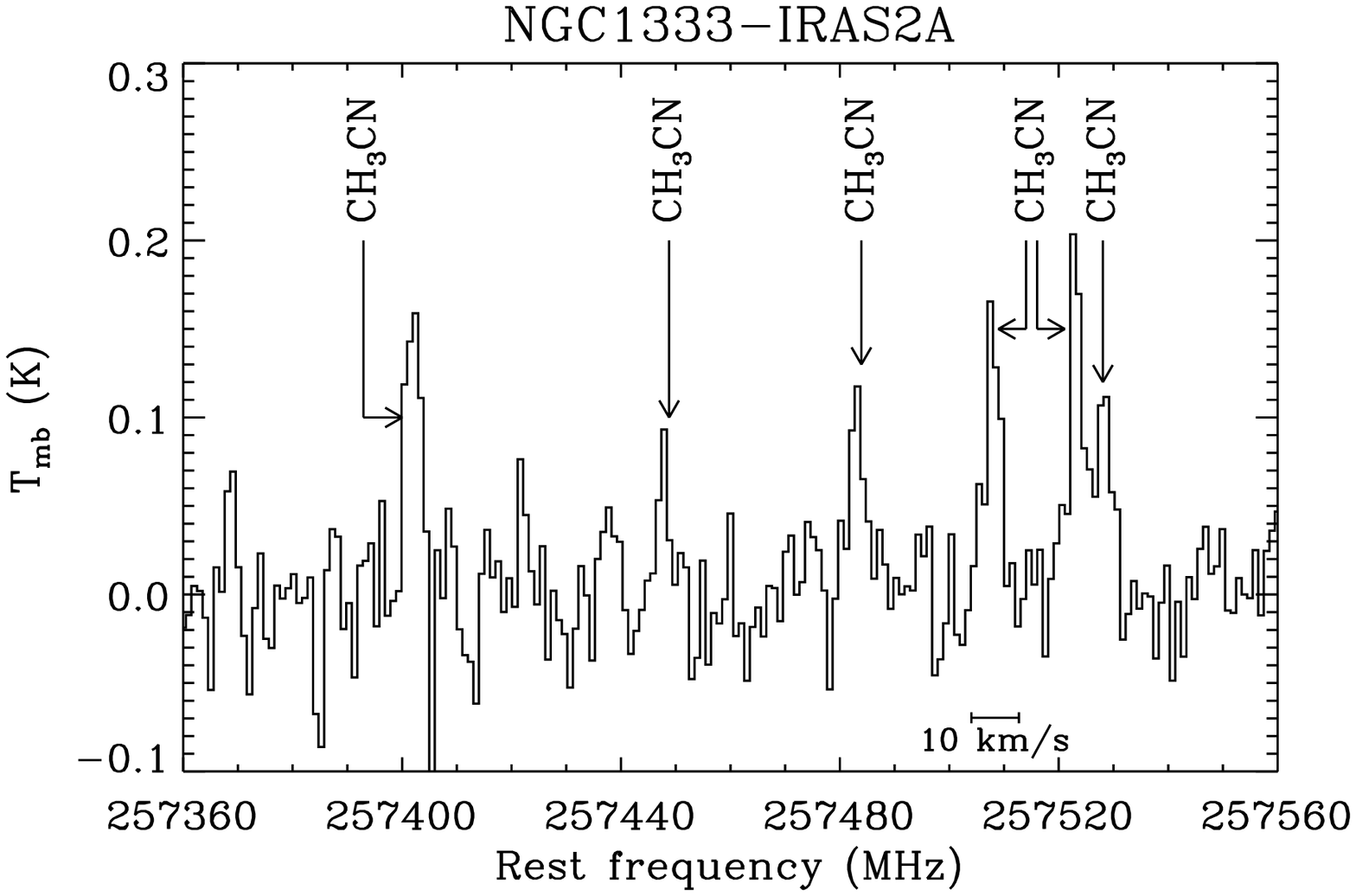}
\caption{Observed spectra towards IRAS4B and IRAS2A.
The rms are 7 mK (top panels) and 5 and 28 mK (bottom panels, left to right);
the spectral resolutions are 0.2, 0.3, 0.8, and 1.2 \kms\ at
90, 98, 110, and 257 GHz, respectively.
Detected transitions are (from left to right, top to bottom):
\mf-E ($8_{0,8}-7_{0,7}$), \mf-A ($8_{0,8}-7_{0,7}$, $7_{2,5}-6_{2,4}$, $8_{3,6}-7_{3,5}$),
\mc\ ($6-4$, $K=4, ..., 0$), and \mc\ ($14-13$, $K=5, ..., 0$).
Unlabeled lines are unidentified.}
\label{spec}
\end{figure*}

Two observation modes were used: position switching with the OFF position at
an offset of $\Delta \alpha$ = --100$''$, $\Delta \delta$ = +300$''$, and 
wobbler switching with a 110$''$ throw in azimuth. Pointing and focus were
regularly checked using planets or strong quasars, providing a pointing 
accuracy of 3$''$. All intensities reported in this paper 
are expressed in units of main-beam brightness temperature. 
At 3, 2, and 1~mm, the angular resolution is 24, 16, and 10$''$ and
the main beam efficiency is 76, 69, and 50\%, respectively.
Figure \ref{spec} shows the obtained spectra.

\section{Results \label{results}}

\begin{table*}
\caption{Molecular lines detected.}
\label{detections}
\begin{center}
\begin{tabular}{lcrrccc}
\hline\hline
Molecule & Transition line & Frequency & E$_u$$^a$   & $T_{\rm mb}$  & $\Delta V ^b$ & $\int T_{\rm mb}$dV \\
         &                 & (MHz)     & (cm$^{-1}$) &      (mK)     & (km s$^{-1}$) &   (K km s$^{-1}$)   \\
\hline
\multicolumn{7}{c}{IRAS4B}\\
\hline
HCOOCH$_3$-A  & $7_{2,5}-6_{2,4}$     &  90156.5  &  13.7 &  26 & 0.6 & 0.017 $\pm$ 0.004 \\
              & $8_{0,8}-7_{0,7}$     &  90229.7  &  13.9 &  23 & 0.5 & 0.013 $\pm$ 0.004 \\
              & $8_{3,6}-7_{3,5}$     &  98611.1  &  18.9 &  21 & 0.7 & 0.016 $\pm$ 0.004 \\
							   	              
HCOOCH$_3$-E  & $8_{0,8}-7_{0,7}$     &  90227.8  &  14.0 &  18 & 1.8$^d$ & 0.035 $\pm$ 0.005 \\
							  	              
CH$_3$CN$^c$						  	              
              & $6_{4,0}-5_{4,0}$     & 110349.7  &  92.3 &  18 & 2.3 & 0.045 $\pm$ 0.009 \\
              & $6_{3,0}-5_{3,0}$     & 110364.6  &  57.6 &  17 & 2.8 & 0.049 $\pm$ 0.011 \\
              & $6_{2,0}-5_{2,0}$     & 110375.1  &  32.8 &  24 & 2.5 & 0.062 $\pm$ 0.013 \\
              & $6_{1,0}-5_{1,0}$     & 110381.5  &  17.9 &  33 & 3.4 & 0.119 $\pm$ 0.014 \\
              & $6_{0,0}-5_{0,0}$     & 110383.6  &  12.9 &  42 & 3.8 & 0.171 $\pm$ 0.015 \\
\hline							  	              
\multicolumn{7}{c}{IRAS2A}\\				  	              
\hline							  	              
CH$_3$CN$^c$						  	              
              & $14_{5,0}-13_{5,0}$   & 257403.6  & 188.5 & 169 & 3.8 & 0.692 $\pm$ 0.091 \\
              & $14_{4,0}-13_{4,0}$   & 257448.9  & 143.9 & 111 & 1.6 & 0.190 $\pm$ 0.046 \\
              & $14_{3,0}-13_{3,0}$   & 257482.7  & 109.1 & 113 & 3.4 & 0.413 $\pm$ 0.044 \\
              & $14_{2,0}-13_{2,0}$   & 257507.9  &  84.3 & 145 & 2.7 & 0.411 $\pm$ 0.055 \\
              & $14_{1,0}-13_{1,0}$   & 257522.5  &  69.4 & 198 & 3.1 & 0.662 $\pm$ 0.087 \\
              & $14_{0,0}-13_{0,0}$   & 257527.4  &  64.4 & 115 & 3.8 & 0.470 $\pm$ 0.068 \\
\hline
\end{tabular}
\end{center}

$^a$ Energy of the upper level of the transition.\\
$^b$ Width of the observed line.\\
$^c$ All the CH$_3$CN lines are unresolved triplets, 
except at 110349.7 MHz, which is an unresolved doublet. The 
quoted signal is the integral over each triplet or doublet.\\
$^d$ This line of the E form of \mf\ has a width about three times as large as that of the 
corresponding transition of the A form, which could be due to the presence of some
unknown transition(s) artificially increasing the width.

\end{table*}

Detected transitions have been identified using the JPL molecular line
catalog \citep{pickett-etal98}, the Cologne Database for Molecular
Spectroscopy \citep{muller-etal01,muller-etal05}, and the National
Institute of Standards and Technology (NIST) Recommended Rest
Frequencies for Observed Interstellar Molecular Microwave Transitions
\citep{lovas04}.  They are reported in Table~\ref{detections}.
We considered only lines with a 3-$\sigma$
detection and a $V_{\rm LSR}$=7.0$\pm$0.4 \kms\
as good identifications.  We detected two of
the five targeted molecules: 4 transitions for HCOOCH$_3$-A in IRAS4B
and 5 and 6 transitions for CH$_3$CN in IRAS4B and IRAS2A,
respectively.  We also have a possible detection for C$_2$H$_5$OH at
226.661 GHz in IRAS4B\footnote{This transition is not seen in
NGC1333-IRAS4A. In this source, there is a tentative detection at
90.118 GHz \citep{bottinelli-etal04-iras4a} that has a lower energy
(3.5 vs 28 cm$^{-1}$), but also lower line strength  
(log(I)=-4.8 vs -3.9).}.
Note that no observations were made at 110 GHz for IRAS2A (see Table
\ref{setups}), and that the rms reached at 257 GHz for IRAS4B is too
high to detect the CH$_3$CN transitions at this frequency, if the line
ratios are similar to those in IRAS4A.\\

\begin{figure*}
\includegraphics[angle=90,width=0.5\textwidth]{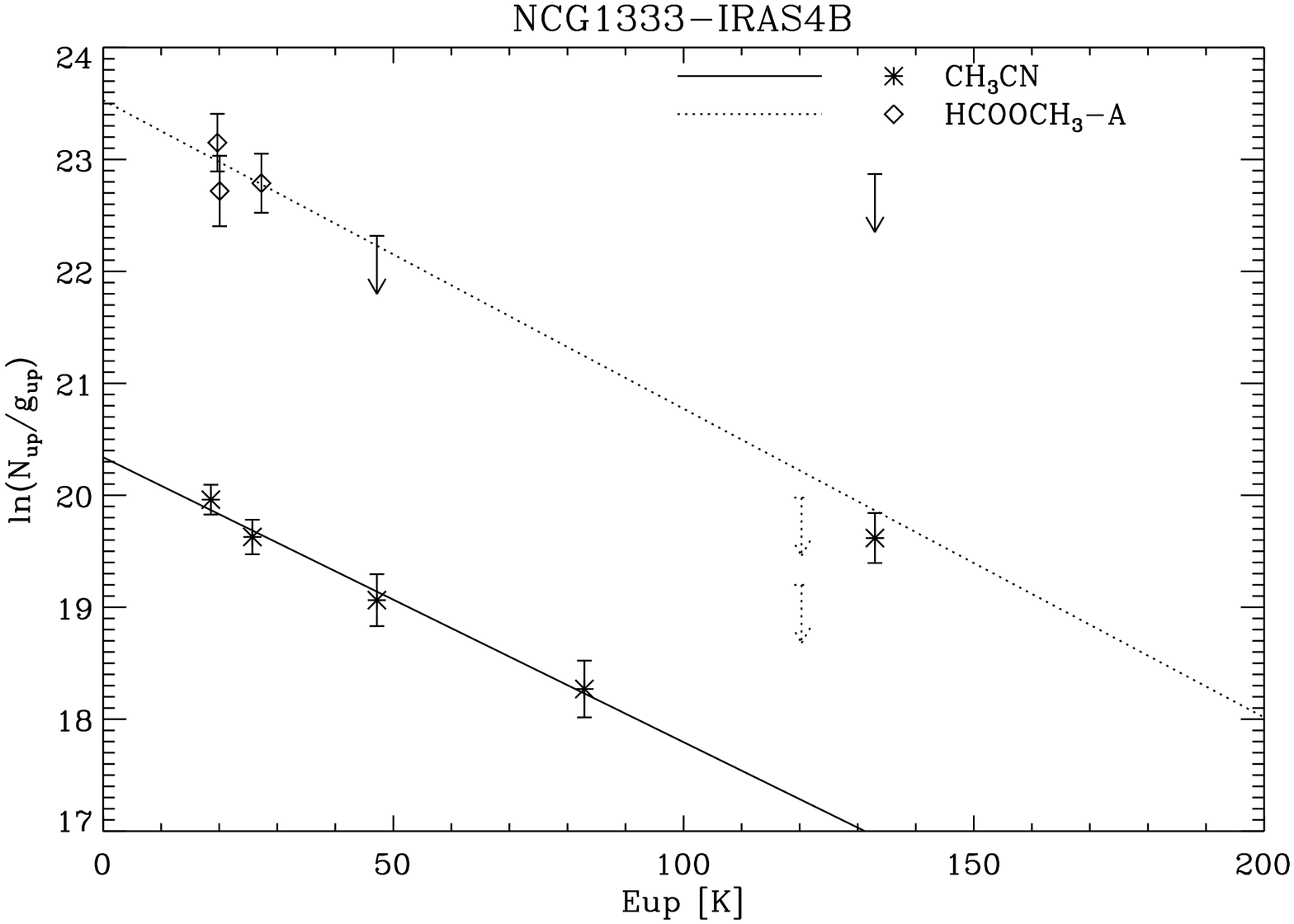}
\includegraphics[angle=90,width=0.5\textwidth]{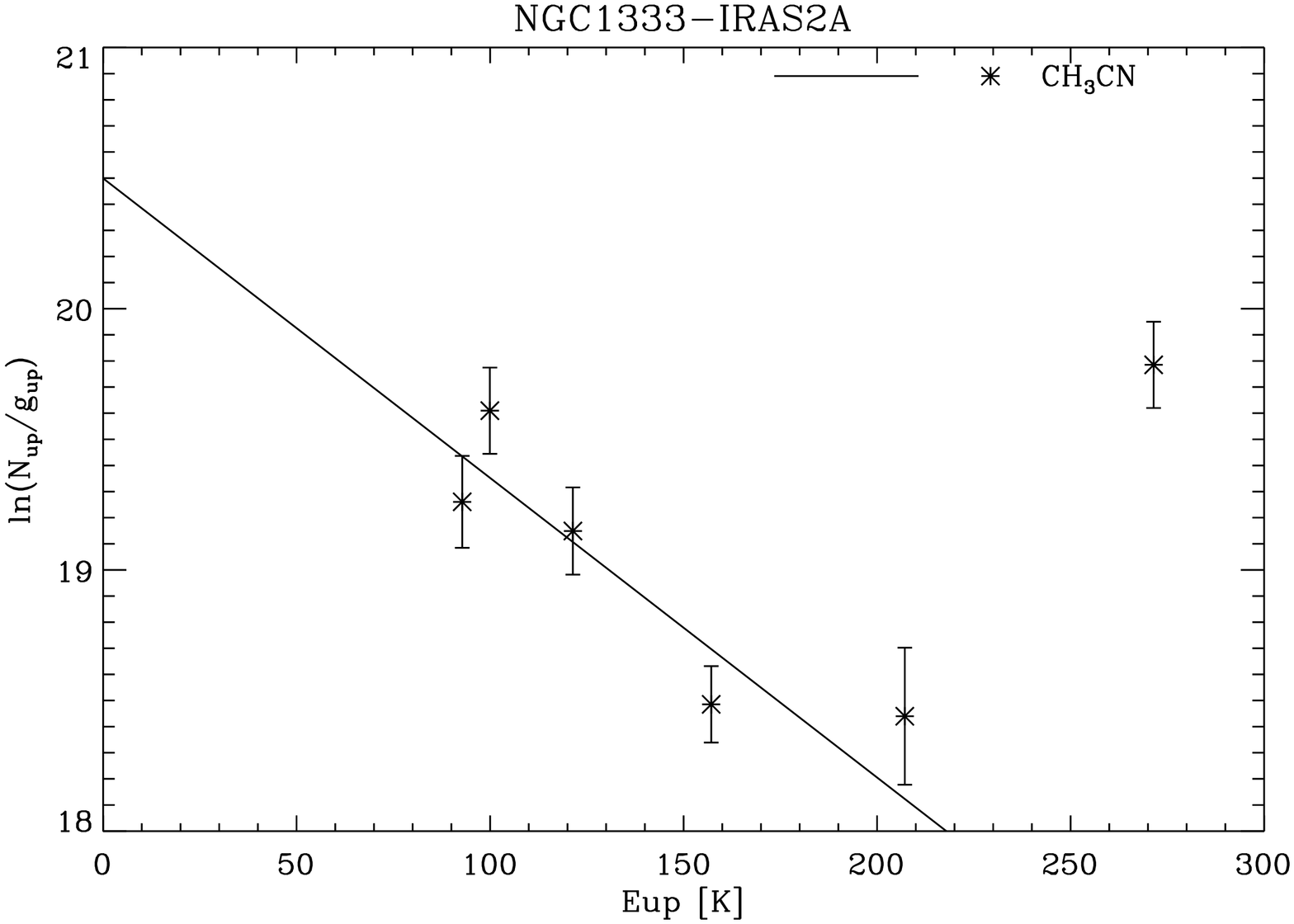}
\caption{Rotational diagrams of the detected molecules, corrected for 
beam dilution at lower frequencies. 
The arrows show the upper limits for undetected transitions.
Lines represent the best fit to the data. Error bars are derived assuming a
calibration uncertainty of 10\% on top of the statistical error. 
The excess of emission of the CH$_3$CN transition at 270 K in IRAS2A 
and 135 K in IRAS4B is probably due to contamination from CH$_3$OH $18_{3,16}-18_{2,17}$
and unknown line(s), respectively, and are not included in the fits.}
\label{rotdia}
\end{figure*}

To derive rotational temperatures and beam-averaged column
densities (see Table \ref{T+N}), we used the rotational diagram method
(Fig.~\ref{rotdia}).  For this analysis, we assumed that the emission from
  a given molecule is unresolved in the smallest beam in which a
  transition of that molecule was detected
($\theta_{\rm min}$, which is always $\geq10''$ 
--- see cols. 3 and 7 in Table \ref{T+N}).
This is justified by the interferometric observations of IRAS16293--2422 
\citep{bottinelli-etal04-iras16293,kuan-etal04,chandler-etal05}, NGC1333-IRAS2A
\citep{jorgensen-etal05-iras2a}, and NGC1333-IRAS4A \citep{bottinelli-etal06},
which show that the bulk of the emission occurs over a region
$\lesssim1''$.
We therefore corrected the integrated intensities
observed at lower frequencies for beam dilution 
with respect to $\theta_{\rm min}$.
Then,
the hot corino abundances were obtained by scaling the beam-averaged
column densities to the sizes of the hot corinos (taken from Table 6
of \citealt{maret-etal04}: 0$\farcs$25 for IRAS4B and 0$\farcs$43 for
IRAS2A) and dividing by the hot corino H$_2$ column
density. 
The latter was obtained from the density profiles derived
by \citet{jorgensen-etal02}, based on 
single-dish observations of the dust continuum
and radiative transfer calculations (see also Table 6 of \citealt{maret-etal04}).
This introduces an uncertainty in the derived abundances, due to the
uncertain sizes and H$_2$ column densities of the hot corinos, caused by 
the extrapolation from single-dish observations. However, abundance ratios are not affected
by this uncertainty, unless different molecules are formed in
different regions (see discussion in Sect. \ref{luminosity}).

We make the following remarks about the results we obtain. The
rotational temperature of HCOOCH$_3$-A is poorly constrained due to
the absence of points at higher energies.
More generally, some of the rotational temperatures are below 100\,K.
This could be due to non-LTE and/or line opacity effects.
We find that the derived
CH$_3$CN abundance is consistent with the value obtained by
\citet{jorgensen-etal05-h2co-ch3oh} for an inner ($T>90$\,K) region.
Upper limits were calculated for the targeted molecules that were not
detected: HCOOH and C$_2$H$_5$CN in both sources, HCOOCH$_3$-A in
IRAS2A, and CH$_3$OCH$_3$ in IRAS4B. The upper limits for the O-bearing
molecules were derived using the rotational temperature derived for
HCOOCH$_3$-A in IRAS4B, whereas the upper limits for N-bearing
molecules were calculated assuming a rotational temperature equal to
that of CH$_3$CN in each source.
Finally, the \dime\ abundance in
IRAS2A was taken from \citet{jorgensen-etal05-iras2a}: the quoted
abundance ($3\times10^{-8}$ in an inner, $T>90$~K region) is
consistent with the upper limit derived from our data
($x<4.2\times10^{-7}$, also calculated using the rotational
temperature derived for HCOOCH$_3$-A in IRAS4B).
\citet{jorgensen-etal05-iras2a} also report a tentative detection of
\mf, but do not give an estimate of the abundance of this molecule.\\

\begin{table*}
\caption{Results from the rotational diagrams and upper limits for IRAS4B and IRAS2A.}
\label{T+N}
\begin{center}
\begin{tabular}{lccccccccc}
\hline\hline
 & \multicolumn{4}{c}{IRAS4B} & & \multicolumn{4}{c}{IRAS2A} \\
\cline{2-5}\cline{7-10}

Molecule & $T_{\rm rot}$ & $\theta_{\rm min}$ & $N_{\rm beam}$$^b$ & $X_{\rm hc}$$^c$ & 
                  & $T_{\rm rot}$ & $\theta_{\rm min}$ & $N_{\rm beam}$$^b$ & $X_{\rm hc}$$^d$ \\
                  &     (K)     &   ($''$)     &      (cm$^{-2}$)      &        & &       (K)     & ($''$)     &  (cm$^{-2}$)     & \\
\hline
HCOOCH$_3$-A  &  38$^{+49}_{-35}$ & 10 & (4.7 $\pm$ 3.6)$\times10^{13}$ & (1.1 $\pm$ 0.8)$\times10^{-6}$ & 
              &  [38]         & 10 & $<2.9\times10^{14}$            & $<6.7\times10^{-7}$ \\
HCOOH         &  [38]         & 22 & $<1.0\times10^{13}$            & $<1.0\times10^{-6}$ &  
              &  [38]         & 25 & $<7.4\times10^{12}$            & $<1.2\times10^{-7}$ \\
CH$_3$OCH$_3$ &  [38]         & 10 & $<6.8\times10^{13}$            & $<1.2\times10^{-6}$ &  
              &  [38]         & 10 & $<1.8\times10^{14}$            & $<4.2\times10^{-7}$ \\
CH$_3$CN      &      39 $\pm$ 3   & 22 & (9.7 $\pm$ 1.0)$\times10^{11}$ & (9.5 $\pm$ 0.2)$\times10^{-8}$ &  
              &      87 $\pm$ 17   & 10 & (3.7 $\pm$ 1.0)$\times10^{12}$ & (8.7 $\pm$ 2.4)$\times10^{-9}$\\
C$_2$H$_5$CN  &  [39]         & 10 & $<4.2\times10^{13}$            & $<7.5\times10^{-7}$ &  
              &  [87]         & 10 & $<4.3\times10^{13}$            & $<1.0\times10^{-7}$ \\
\hline
\end{tabular}
\end{center}
{\sc Note} --- When they could not be derived from a rotational diagram, 
temperatures have been assumed to be equal to that of \mf\ and \mc\ 
for O-bearing and N-bearing molecules, respectively. 
In this case, the values are shown in square brackets. \\
$^a$ Smallest beam size for which a transition was detected 
(see text for details).\\
$^b$ Column density averaged over $\theta_{\rm min}$.\\
$^c$ Abundance in the hot corino assuming an H$_2$ column density 
of $N$(H$_2$) = 8.1 $\times$ 10$^{22}$ cm$^{-2}$ and a hot corino size 
of 54 AU, i.e., 0.25$''$ at 220pc \citep[from][]{maret-etal04}.\\
$^d$ Abundance in the hot corino assuming an H$_2$ column density 
of $N$(H$_2$) = 2.1 $\times$ 10$^{23}$ cm$^{-2}$ and a hot corino size 
of 94 AU, i.e., 0.43$''$ at 220pc \citep[from][]{maret-etal04}.\\
\end{table*}

%%%%%%%%%%%%%%%%%%%%%%%%%%%%%%%%%%%%%%%%%%%%%%%%%%

\section{Discussion \label{discussion}}

To investigate whether the data provide information on the
formation of complex molecules, it is necessary to know what the
possible formation paths are. 
We therefore report the formation reactions found in the
literature for the detected molecules in Appendix 
\ref{formation}, and we bring up some key points
regarding these formation routes in Sect. \ref{summary-formation}.
Since low-mass protostars were
thought to have insufficient luminosity to develop a hot core-type
region, we will then investigate the potential dependence of the
complex molecules' abundances on the luminosity (Sect. \ref{luminosity}).  In Sect.
\ref{abundance-ratios}, we will analyze the hot corino data in view of
the information given in Appendix \ref{formation} and look at the
differences with their massive counterparts and with Galactic center
clouds in Sects. \ref{comparison} and \ref{gc}, respectively.

\subsection{Notes on the formation routes of complex organic molecules 
\label{summary-formation}}

For all the complex molecules considered here, we can see from
Appendix \ref{formation} that grain-surface formation
is a possible alternative to the ``classical'' gas-phase formation.
In this classical view, an estimate of the formation and 
destruction timescales for \dime\ indicates that destruction of this
molecule is likely compensated by formation mechanisms (Sect. \ref{timescales}).
Regarding grain-surfaces processes, it is
usually implicitly assumed that they are taking place
during the cold phase preceding the warm-up of the dust by the newly
born star. However, as the protostar heats up, it is improbable that
the dust temperature suddenly jumps from $\sim$10 to
$\sim$100~K. Instead,
it is more likely that the temperature ``slowly''
rises across the inner envelope, leading to a gradual heating of the grains
(e.g., \citealt{viti-etal04,garrod+herbst06}). The effect of this
gradual temperature change on grain-surface and gas-phase chemistry
can be quite substantial. Indeed, the modeling by
\citet{garrod+herbst06} shows that gas-phase and grain-surface
chemistries are strongly coupled during the warm-up phase: molecules
formed on the grain can evaporate and affect the gas-phase chemistry,
whose products can re-accrete and in turn change the grain-surface
chemistry.
\\ 

Note that for some molecules, grain-surface formation is not only an
alternative, but also apparently the only choice.  This is the case
for \mf\ and even more so for HCOOH since it has been detected in
the ices of star forming regions
\citep[e.g.,][]{schutte-etal99,keane-etal01}, as well as of quiescent
molecular clouds \citep{knez-etal05}.  In particular, the detection of
icy HCOOH in quiescent clouds would indicate not only that HCOOH
formation occurs on the grain surface, but also that it would pre-date
the first phase of the star formation mechanism, supporting the theory
of formation in the cold, rather than warm-up phase.
Whether
the other complex organic molecules could also follow the same pattern 
cannot be commented upon since, to our knowledge, no
other complex molecule has been detected in ices.
This is because, unfortunately, the infrared spectra of complex
organic molecules are not well known in this medium.
The difficulty in determining the presence of these molecules lies in the
fact that they may produce only slight shifts and broadenings, and
that the peak positions are characteristic of functional groups, not
molecular species (Tielens, priv. comm.). Hence, specific
identifications are always somewhat ambiguous. 
Overall, it also stands out from Appendix \ref{formation} that more
experiments on thermal surface chemistry at low temperatures are
needed to check the feasibility of the outlined reactions. 
Cosmic ray processing of icy grain mantles may also be able to produce
complex molecules, but this process requires more
quantitative modeling to compare with observations.\\

\subsection{Luminosity dependence \label{luminosity}}

As mentioned previously, before the discovery of a hot corino around
IRAS16293, it was believed impossible to have chemically rich regions
driven by gas-phase reactions
following mantle evaporation around low-mass protostars. The argument
was that, given the low luminosity of these objects, the regions where
the ices sublimate would be so small that the gas crossing-time would
be shorter than the time needed to form complex organic molecules in
the gas phase \citep[e.g.,][]{schoier-etal02}.  This is in fact not the
case since several hot corinos have now been discovered \citep[][ this
work]{cazaux-etal03,bottinelli-etal04-iras4a,jorgensen-etal05-iras2a}.
However, the question remains regarding the impact of the luminosity on the
abundances of complex organic molecules in hot corinos.\\  

\begin{figure*}
\includegraphics[bb=132 16 499 774,height=\textwidth,angle=90]{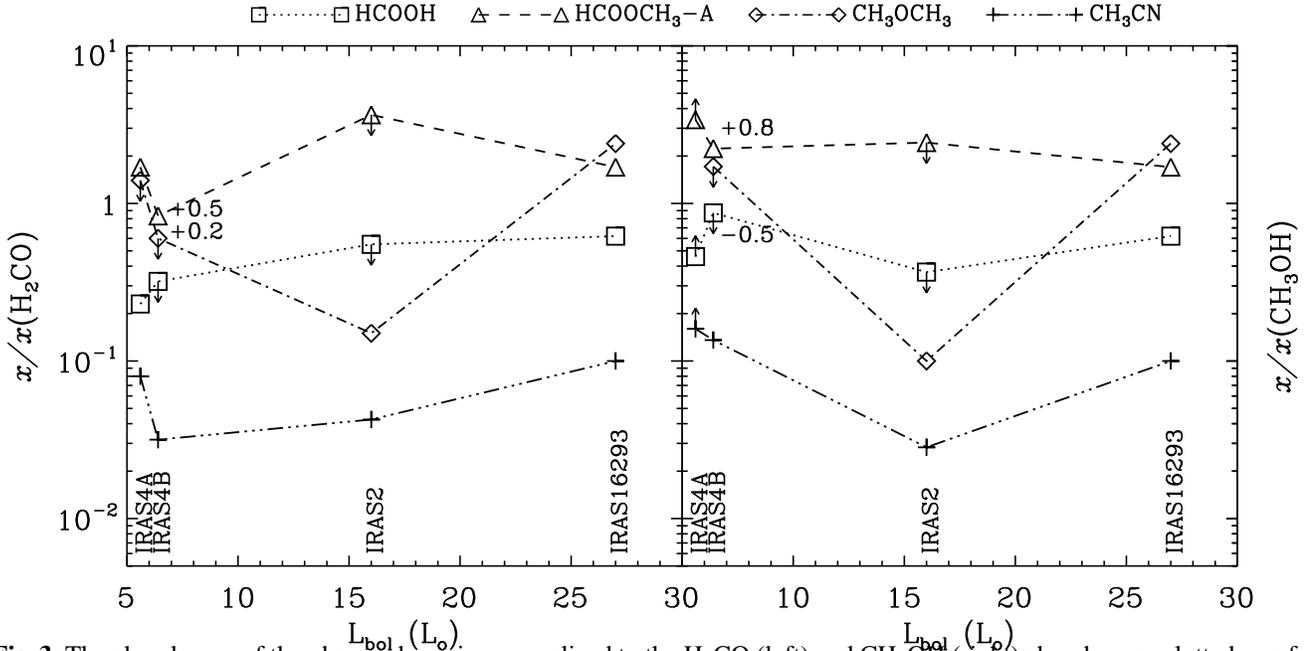}
\caption{The abundances of the observed species normalized to the
H$_2$CO (left) and CH$_3$OH (right) abundances, plotted as a function
of bolometric luminosity.  Square, triangles, diamonds, and plus signs
represent HCOOH, \mf, \dime, and \mc, respectively.  The abundance for
\dime\ in IRAS2A was taken from \citet{jorgensen-etal05-iras2a} and is
likely underestimated (see text). Note that we only have an upper
limit on the CH$_3$OH abundance in the hot corino of IRAS4A,
therefore, we did not plot the point corresponding to CH$_3$OCH$_3$
for which only an upper limit is available in that source. Also, we do
not show \ec\ since only upper limits are available in three of the
four sources.  }
\label{xratio}
\end{figure*}

Table
\ref{abundances} lists the measured abundances and upper limits for
the four hot corinos. To remove the uncertainty on the sizes
of the hot corinos, we choose to look at abundance ratios, in
particular with respect to formaldehyde and methanol since these
molecules have been proposed to be the parent molecules for complex
oxygen-bearing species, if they are formed in the gas-phase.  These
abundance ratios are plotted on Fig. \ref{xratio} as a function of
the bolometric luminosity of the low-mass protostars.  For
information, the abundance ratio for \mc\ is also plotted (\ec\ was
not included due to the number of upper limits), although formaldehyde
and methanol are not thought to be the parent molecules of
nitrogen-bearing species (see Sect. \ref{n-bearing}).  Note that the
abundance ratios of \dime\ with respect to both \fdh\ and \m\ in
IRAS2A seem to be ``outliers'' compared to the other protostars and to
other O-bearing molecules. However, recall that the \dime\ abundance
was taken from \citet{jorgensen-etal05-iras2a} where it was derived
from only one detected transition.  Apart from the \dime\ points and
taking into account the uncertainties pertaining to abundance
determination, we can see from Fig. \ref{xratio} that the abundance
ratios of complex molecules with respect to \fdh\ or \m\ do not depend
on the luminosity, in the range $\sim 5-30$~L$_\odot$.  Since the
abundances of \fdh\ or \m\ are not themselves a function of luminosity
\citep[see][]{maret-etal04,maret-etal05}, then {\it the absolute
abundances of the complex species do not depend on the luminosity of
the protostar}.  
Whatever the formation mechanism, either in the
gas phase or on the grain surfaces, the efficiency in forming complex
organic molecules is largely constant in the range of studied
luminosities. Since the luminosity, together with the density, defines
the radius at which ices sublimate, this also implies that this
efficiency is rather constant in the inner 200 AUs or so of the
studied sources, despite the different involved densities (from 10$^6$
to 10$^{9}$ cm$^{-3}$, \citealt{maret-etal04}). \\

We also investigated the possible dependence of the abundance ratios
on the ratio of submillimeter to bolometric luminosity, L$_{\rm
smm}$/L$_{\rm bol}$, since it has been suggested as an indicator of
evolutionary stage. We find that the abundance ratios do not depend on
this parameter either. \citet{maret-etal04} found an apparent
anti-correlation between the inner abundance of \fdh\ and L$_{\rm
smm}$/L$_{\rm bol}$ and proposed that this could be explained if
L$_{\rm smm}$/L$_{\rm bol}$ depends on the initial conditions of the
protostars rather than their evolutionary stage.  Indeed, more atomic
hydrogen is available in less dense (i.e., with a higher L$_{\rm
smm}$/L$_{\rm bol}$) environments, which leads to the formation of
more \fdh\ and \m.  If we plot the inner \m\ abundance as a function
of L$_{\rm smm}$/L$_{\rm bol}$, we also notice an apparent
anti-correlation. Since the abundance ratios are roughly constant with
different L$_{\rm smm}$/L$_{\rm bol}$, then the absolute abundances of
complex molecules should also be anti-correlated with this parameter.
Following a similar line of thought as \citet{maret-etal04} and
assuming that complex O-bearing molecules form on grain surfaces via
H, O, OH, and/or CH$_3$ additions, this anti-correlation could be
indicative of these species being more readily available in less dense
environments. \\

\begin{table*}
\caption{Abundances of parent and daughter molecules in our sample of four
low-mass protostars.}
\label{abundances}
\begin{center}
\begin{tabular}{lccccc}
\hline\hline
Molecule & IRAS16293     & IRAS4A    & IRAS4B & IRAS2A & Ref.\\
\hline
H$_2$CO         & 1$\times10^{-7}$     &   2$\times10^{-8}$ &    3$\times10^{-6}$ & 2$\times10^{-7}$    & 1,2,3 \\
CH$_3$OH        & 1$\times10^{-7}$     &  $<1\times10^{-8}$ &    7$\times10^{-7}$ & 3$\times10^{-7}$    & 4 \\
HCOOCH$_3$-A$^a$   & 1.7$\times10^{-7}$   & 3.4$\times10^{-8}$ &  1.1$\times10^{-6}$ & $<6.7\times10^{-7}$ & 5,2,6 \\
HCOOH           & 6.2$\times10^{-8}$   & 4.6$\times10^{-9}$ & $<1.0\times10^{-6}$ & $<1.2\times10^{-7}$ & 5,2,6 \\
CH$_3$OCH$_3$   & 2.4$\times10^{-7}$   & $<2.8\times10^{-8}$& $<1.2\times10^{-6}$ & 3.0$\times10^{-8}$  & 5,2,6,7 \\
CH$_3$CN        & 1.0$\times10^{-8}$   & 1.6$\times10^{-9}$ &  9.5$\times10^{-8}$ & 8.7$\times10^{-9}$  & 5,2,6 \\
C$_2$H$_5$CN    & 1.2$\times10^{-8}$   & $<1.2\times10^{-9}$& $<7.5\times10^{-7}$ & $<1.0\times10^{-7}$ & 5,2,6 \\
\hline
\end{tabular}
\end{center}
{\sc Note} --- Except for \fdh\ and \m, for which \citet{maret-etal04,maret-etal05} took into
account the effects of opacity to derive the abundances of these species, it was not possible
to determine the optical thickness of the complex molecules' transitions, so that all the complex
molecules' abundances should be considered as lower limits.\\
$^a$ We report here the abundances of the A form only of \mf, for which we have the
highest number of rotational diagrams. Note that the abundance of the E form is usually
very close to that of the A form (from IRAS16293 and IRAS4A; 
\citealt{cazaux-etal03,bottinelli-etal04-iras4a}), so that the total \mf\ abundance
would be twice that of the A form.\\
{\it References} -- (1) \citet{ceccarelli-etal00-h2co16293}. (2) \citet{bottinelli-etal04-iras4a}.
(3) \citet{maret-etal04}. (4) \citet{maret-etal05}. (5) \citet{cazaux-etal03}.
(6) This work. (7) \citet{jorgensen-etal05-iras2a}. 
\end{table*}

\subsection{Dependence on methanol and formaldehyde hot corino abundances \label{abundance-ratios}}

\begin{figure*}
\includegraphics[bb=320 15 590 775,height=\textwidth,angle=90]{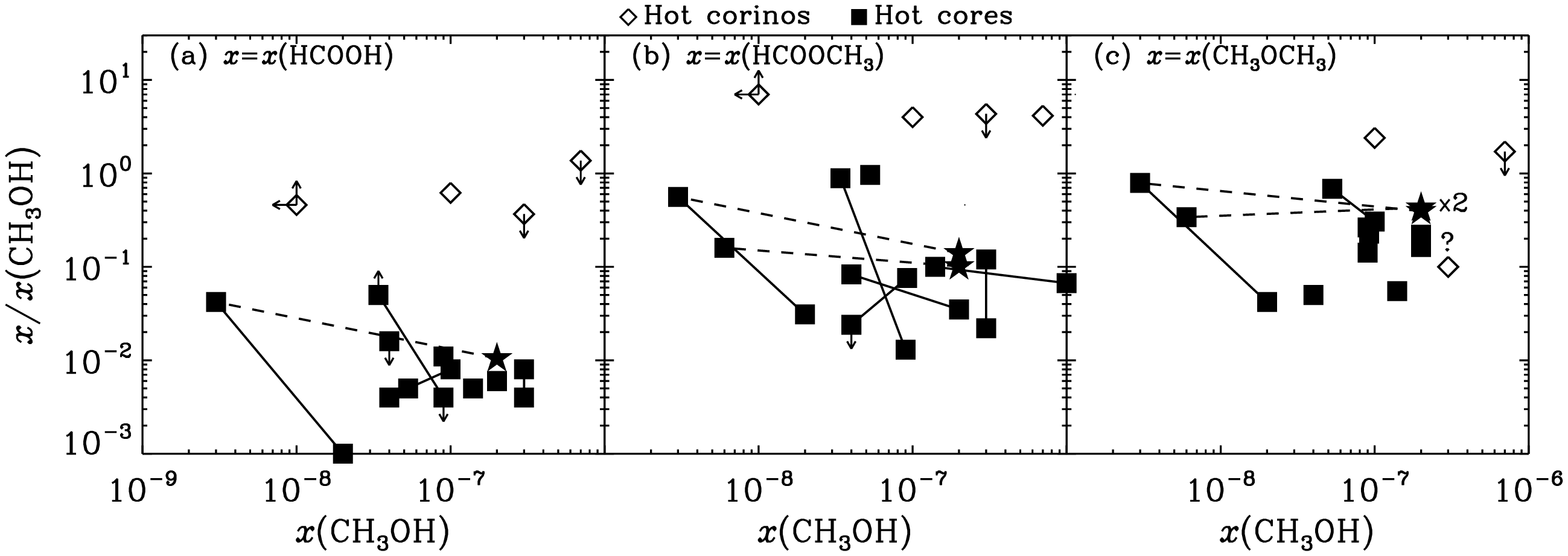}
\caption{Abundance ratios of complex O-bearing molecules to methanol,
plotted as a function of the methanol abundance. Open diamonds
represent hot corinos. Filled squares represent abundances ratios of
hot cores, derived from beam-averaged column density analysis. Squares
linked by a solid line represent abundance determinations from
different authors (\citealt{gibb-etal00-g327} and references therein,
\citealt{ikeda-etal01}).  Stars represent the hot cores of SgrB2 (N)
and (M), where an analysis of the methanol emission similar to what
done in the hot corinos has been carried out
\citep{nummelin-etal00}. The dotted lines connect the SgrB2 (N) and
(M) hot cores (stars) to the squares corresponding to the
cold envelopes of these sources. See the text for further
details. Note that panel (b) represents the total \mf\ abundances,
that is twice the abundances of the A form quoted in Table \ref{abundances}.
The point corresponding to \dime\ in IRAS4A is not
represented due to both \dime\ and \m\ abundances being upper limits
in this source. The question mark refers to the ratio in IRAS2A 
and indicates that the \dime\ abundance is likely underestimated 
in this source (see Sect. \ref{luminosity}).}
\label{ratio-ch3oh-all}
\end{figure*}

\begin{figure*}
\includegraphics[bb=320 15 590 775,height=\textwidth,angle=90]{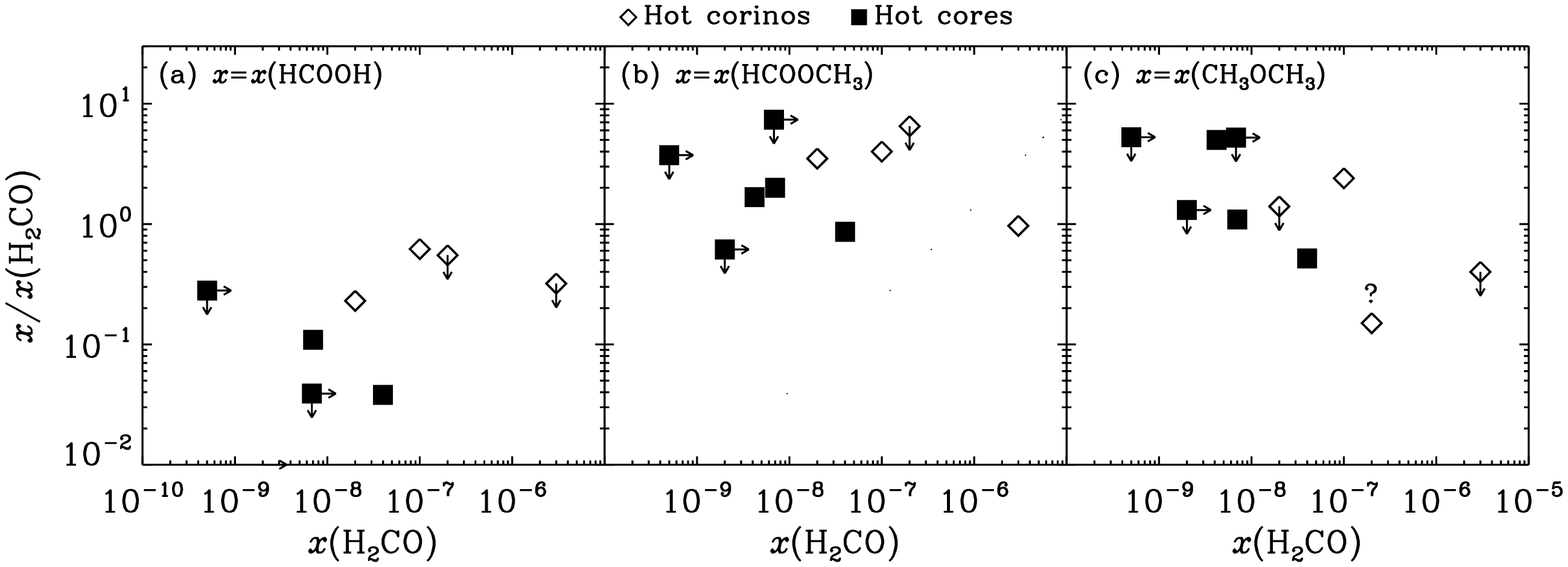}
\caption{Abundance ratios of complex O-bearing molecules to
formaldehyde, plotted as a function of the formaldehyde
abundance. Open diamonds
represent hot corinos. Filled squares represent abundances ratios of hot
cores, derived by beam-averaged column density analysis. 
}
\label{ratio-h2co-all}
\end{figure*}

Whether we consider gas-phase or grain-surface formation, \m\ and
\fdh\ appear as key molecules: in the first case, they have been
suggested as parent molecules, in the second, they are known mantle
constituents. It is therefore interesting to investigate the abundance
ratios of complex molecules to \m\ and \fdh\ as a function of \m\ and
\fdh\ abundances themselves. We plot these quantities for hot corinos
in Figs. \ref{ratio-ch3oh-all} and \ref{ratio-h2co-all},
respectively. From these figures, we can see that:
\begin{itemize}
\item[(i)] The complex molecules in hot corinos have comparable abundance ratios, apparently
 independent of the \m\ and \fdh\ abundances. 
\item[(ii)] These abundance ratios are close to unity.
\end{itemize}
The implications are:
\begin{itemize}
\item {\it In the case of gas-phase formation from methanol or
formaldehyde:} (i) and (ii) mean that, in all the hot corinos, the
formation of complex molecules uses up a significant fraction of the
parent molecules sublimated from the mantles. However, gas-phase
chemistry does not seem able to reproduce this behavior.  For
example, the collapsing envelope model of
\citeauthor{rodgers+charnley03} (\citeyear{rodgers+charnley03}; see
also \citealt{rodgers+charnley01}) predicts \dime\ to \m\ abundance
ratios of only $10^{-2}-10^{-1}$. Similarly,
\citet{rodgers+charnley01} predicted \mf\ to \m\ abundance ratios
$<7\times10^{-3}$, and we now know that they used too high a formation
rate coefficient for \mf, so that the actual prediction should be even
smaller \citep{horn-etal04}.  Therefore, either gas-phase models are
not adequate, or complex molecules are not formed in the gas phase.

\item {\it In the case of grain-surface formation:} (i) and (ii) show
that complex molecules are as important mantle constituent as \m\ and
\fdh. Observations of solid HCOOH and \m\ along quiescent lines
of sight by \citet{knez-etal05} and in protostars \citep{keane-etal01}
support this idea since the quoted HCOOH and \m\ abundances yield
abundance ratios of order unity, as we find for the hot corinos. No
observations of other complex molecules in the ices are available, but
their presence cannot be excluded considering, as mentioned in Sect.
\ref{summary-formation}, the difficulty of identifying their signature
in infrared spectra.
\end{itemize}

Note that gas-phase reactions proposed in Sect.
\ref{hcooh-formation} for the formation of HCOOH do not involve \m\ or
\fdh. Therefore the conclusions mentioned for gas-phase formation do
not apply to this molecule, whereas the analysis regarding
grain-surface formation is still valid.\\ 
Using the methanol abundances derived by
\citet{jorgensen-etal05-h2co-ch3oh} does not change the shape of
Fig. \ref{ratio-ch3oh-all} since their values are comparable to the
ones derived by \citet{maret-etal05}. However, if we take formaldehyde
abundances from \citet{jorgensen-etal05-h2co-ch3oh}, we notice that
the abundance ratios are larger than those plotted in Fig.
\ref{ratio-h2co-all} (as expected since
\citealt{jorgensen-etal05-h2co-ch3oh} do not model any \fdh\
abundance jump), and that the abundance ratios are scattered by up to
two orders of magnitude.\\
Overall, in the \citet{maret-etal04,maret-etal05} framework, our data
are consistent with either gas-phase or grain-surface formation of
complex molecules in hot corinos, with nonetheless more support for
the later route.  However, other scenarios could be considered. For
example, complex molecules could form in the ISM, deplete onto
grain mantles during the accretion phase and desorb as the protostar
heats up its environment.  If this formation in the ISM were true, we
should be able to observe their low-energy transitions in dark clouds
or in the cold envelopes of the protostars.  For example,
\citet{remijan+hollis06} observed a transition of HCOOH with E$_u$=3.2
K around IRAS16293B for which the spatial distribution is around (and
not peaking at) the position determined from the continuum.  We cannot
therefore rule out this theory, but more observations are needed to
support it.

\subsection{Comparison hot corinos -- hot cores \label{comparison}}

Due to the physical differences between hot corinos and hot cores,
comparing these two types of objects can potentially tell us a lot
about the parameters influencing the formation and evolution of
complex molecules. We therefore searched the literature for the
relevant abundances in hot cores to calculate abundance
ratios in these objects as well.  To our knowledge, there are only
four hot cores (Sgr B2(N), Orion hot core, Orion
compact ridge, and W3(H$_2$O)) for which data on all five O-bearing molecules studied
here (\fdh, \m, HCOOH, \mf, and \dime) are available, but we can find
another seven sources (Sgr B2(M), G327.30$-$0.60, G34.26$+$0.15,
NGC6334(I), W51 e1/e2, G31.41$+$0.31, and G10.47$+$0.03) for which \m\
and at least two of the complex molecules molecules (HCOOH, \mf, and
\dime) have measured abundances. Data are taken from
\citet{gibb-etal00-g327} and references therein, and from
\citet{ikeda-etal01}. 

Figures \ref{ratio-ch3oh-all} and
\ref{ratio-h2co-all} plot the abundance ratios of massive hot cores,
as derived from the literature. The first thing to notice is that the
abundance ratios in the massive hot cores have been derived by
beam-averaged column densities, except for SgrB2 (N) and (M), where
\citet{nummelin-etal00} have carried out an analysis of the methanol
emission similar to what has been done for the hot corinos, namely
disentangling the hot core from the cold outer envelope contribution
in the beam.  
For a given complex molecule, if a source was listed in both
\citet{gibb-etal00-g327} and \citet{ikeda-etal01}, 
then two abundance ratios were derived from the
different authors. In this case, the two values are reported in
in Fig. \ref{ratio-ch3oh-all} and linked with a solid line.
Disturbingly, they can differ by almost two
orders of magnitude, and the cause of
this discrepancy is unclear.
In Fig. \ref{ratio-ch3oh-all}, we
also report the results obtained from the hot core + outer envelope
analysis of CH$_3$OH emission in SgrB2 (N) and (M), to show the
uncertainty associated with the different methods of abundance ratios'
determinations. As can be seen, there is only a factor 4
 difference in the abundance ratios if the hot core or the outer
envelope abundance of methanol is used.\\

In principle, the different methods used to derive the abundance
ratios in hot cores (beam-averaged column density) and corinos (full
density and chemical structure analysis) may lead to such a different
result that the comparison may be meaningless. However, while this
would certainly be the case for the absolute values of the abundances,
the abundance ratios suffer much less from the different 
methods, as shown by the ``small'' factor 4 in the SgrB2 (N) and (M)
sources \citep{nummelin-etal00}. In practice, the beam-averaged
abundance ratios are very different from the reality only if the
spatial distributions of the different molecules in the considered
source are different, and/or if there is a large contribution from the
outer envelope with respect to the hot core. While a direct measure of
the molecular emission extent would require carrying out interferometric
observations, the available data can give
a good hint on where the emission originates from, by looking at the
rotational temperature, $T_{\rm rot}$.  As a matter of fact, the
$T_{\rm rot}$ of 
the hot cores considered in this study (with the possible exception of
the H$_2$CO and HCOOH) are
implying that the emission is indeed dominated by the hot cores rather
than the cold envelope (which may also be due to an observational
bias, that is, if the observed transitions are probing warm rather than
cold gas).  In this respect, therefore, we think that the comparison
between hot corinos and hot cores shown in Fig.
\ref{ratio-ch3oh-all} is reliable. There is, however, more uncertainty
associated with the hot cores' abundance ratios reported in Fig.
\ref{ratio-h2co-all}, since H$_2$CO emission could have an important
contribution from the cold envelope. \\

The abundance ratios with respect to \m\ and \fdh\ were noticed to be
roughly constant for hot corinos. Regarding hot cores, the ratios seem
to decrease with increasing \m\ or \fdh\ abundance, but the data are
also consistent with a constant ratio with a larger scatter.  In any
case, it is clear from these figures that {\it the abundance ratios
with respect to methanol in hot cores are lower than in hot corinos}
by 1--2 orders of magnitude, whereas abundance ratios with respect to
formaldehyde are comparable in hot corinos and hot cores. We also
notice that the HCOOH abundance ratios in hot cores are about one
order of magnitude lower than the \mf\ and \dime\ abundance ratios,
whereas they are lower by only about a factor four in hot corinos.
For completeness, we considered how Figs. \ref{ratio-ch3oh-all}
and \ref{ratio-h2co-all} would change if we were to use beam-averaged
abundances ratios in hot corinos, as in hot cores. In this case,
abundance ratios in hot corinos would be smaller by a factor
$\lesssim10$: ratios for \mf\ and \dime\ would become comparable in
the two types of objects, while HCOOH ratios would still be larger in
hot corinos than in hot cores. This is very likely due to the relative
larger contribution of methanol and formaldehyde emission in the cold
envelope in low mass with respect to high mass protostars. Indeed,
high energy transitions are more easily detected in high than low mass
protostars \citep[e.g.,][]{comito-etal05,schilke-etal01,schilke-etal97,blake-etal95}, 
which support the above interpretation.\\

Whether considering beam-averaged ratios in all objects or not,
Fig. \ref{ratio-ch3oh-all} (and to a lesser extent Fig. \ref{ratio-h2co-all}) show that
{\it hot corinos are not just scaled versions of hot cores} and that
in fact, complex molecules are relatively more abundant in hot corinos
than in hot cores.  This conclusion would still be valid if we assume
\m\ and \fdh\ abundances from \citet{jorgensen-etal05-h2co-ch3oh}
since, as mentioned in the previous section, the abundance ratios with
respect to \m\ and \fdh\ would be similar and higher, respectively, so
in any case, they would be higher than the abundance ratios 
in massive hot cores.
A possible explanation for the difference between hot cores and hot
corinos could lie in the grain mantle composition. Indeed, we already
know that the levels of deuteration differ in these two types of
objects, more specifically that extreme deuteration occurs in low-mass
but not in high-mass protostars (e.g., \citealt{ceccarelli-etal06}, and
references therein).  Moreover, \citet{boogert-etal04} summarize the
abundances of some mantle constituents (H$_2$O, CO, CO$_2$, \m,
OCN$^-$) in two low-mass and high-mass embedded protostars, and 
large differences can also be seen there. It would therefore not be
surprising that this would also be the case for complex molecules if
they formed on the grain surfaces.  The reason could be, for example,
that hot corinos are preceded by a longer cold phase during which
grain-surface reactions are at play, so that complex molecules in hot
cores would not have time to become an important grain-mantle
constituent. \\ Alternatively, the difference in abundance ratios
could be due to a difference in the chemistry, since this depends on
environmental parameters such as the density and temperature of the
gas.  For example, the model of \citet{rodgers+charnley01} shows that
the predicted abundances are different at 100 or 300~K. Also,
considering for example the methyl formate, the gas-grain model in the
warm-up phase presented in \citet{garrod+herbst06} shows that the
gas-phase formation of this molecule is more efficient at low
temperatures, and that the grain-surface pathway leads to higher
abundances the longer the dust temperature remains in the $40-60$~K
range. Either way, formation of complex molecules during the warm-up
stage of the protostar would be more efficient in low- than in
high-mass environments.  However, observations performed towards
Galactic center clouds (see Sect. \ref{gc}) are acting against this
theory since complex molecules are observed in these clouds although
they are not subject to a warm-up phase.\\  Finally, recall that all
the proposed grain-surface formation routes of complex molecules
involved UV or cosmic-ray processing. Hot cores/corinos are shielded
from external UV radiation fields, but low-mass protostars, unlike
massive ones, are known to be powerful X-ray sources \citep[e.g.,][]
{feigelson+montmerle99}. X-rays have already been proposed to be the
reason for the presence of calcite in the low-mass protostars
\citep{chiavassa-etal05, ceccarelli-etal02-calcite}.  In our case,
X-rays emitted by Class 0 objects could provide the necessary energy
to produce a large number of radicals and hence a large amount of
complex molecules on the grain surfaces, thereby explaining the larger
abundance ratios observed in hot corinos compared to hot cores.  \\

\subsection{Comparison with Galactic center clouds \label{gc}}

Hot cores/corinos are not the only objects where complex molecules have been observed. Indeed, \citet{requena-torres-etal06} have carried out a survey of complex O-bearing molecules in Galactic center (GC) clouds. 
These clouds are known to possess a warm ($>100$~K) and not too dense 
($\sim10^4$~\cc) gas \citep{huettemeister-etal93,rodriguez-fernandez-etal00,
ceccarelli-etal02-sgrb2}. This warm component is probably caused by shocks \citep[e.g.,][]{flower-etal95,rodriguez-fernandez-etal04} arising from cloud-cloud collisions \citep{huettemeister-etal93,huettemeister-etal98}.
These shocks are thought to be at the origin of the enhanced \ammonia\ \citep{flower-etal95},
SiO \citep{martin-pintado-etal97}, and C$_2$H$_5$OH \citep{martin-pintado-etal01} 
abundances.\\
Surprisingly enough, considering the very different physical environments between GC clouds
and hot corinos, 
\citet{requena-torres-etal06} found, as we do in the present work for hot corinos, that the ratios of the abundances with respect to \m\ (which are comparable to those reported in Fig. \ref{ratio-ch3oh-all} for hot cores) are approximately constant and do not depend on the \m\ abundance. 
The authors conclude that the gas-phase abundances 
of the organic molecules they observed in GC clouds
are likely due to the formation of these molecules on the grain surfaces and their release in the gas-phase 
from sputtering/erosion of the grain mantles by the shocks, as it is the case for ammonia, silicon oxide, and ethanol.\\

Note that there is evidence for the presence of X-rays in these GC clouds \citep{martin-pintado-etal00}. These authors suggest that X-rays could contribute to the formation of molecules on grain
surfaces and evaporate small dust grains. This theory adds some support to the possible role played by
X-rays in the formation of complex organic molecules.

\section{Conclusion \label{conclusion}}

In this paper, we presented the detections of methyl formate and/or methyl cyanide in the low-mass protostars IRAS4B and IRAS2A, confirming the presence of a hot corino in their inner envelope. 
The conclusions arising from the analysis of these observations combined with data on 
the two other hot corinos (IRAS16293 and IRAS4A) are:
\begin{itemize}
\item {\it Hot corinos are a common phase in the formation of
solar-type protostars} and complex organic molecules are ubiquitous in Class 0 protostars.
\item The absolute abundances (i.e., with respect to H$_2$)
of complex molecules in hot corinos do not depend on the 
bolometric luminosity.
available in less dense environments.
\item Abundance ratios of complex molecules' abundances to \m\ or \fdh\ abundances
($x/x(\m)$ and $x/x(\fdh)$
are of order unity and do not depend on \m\ or \fdh\ abundances, 
indicating that complex molecules form on grain surfaces or that gas-phase models have to be
revised.
\end{itemize}

Furthermore, we compared hot corinos with massive hot cores.
Keeping in mind the different methods of abundance determination used in the two types
of objects, we found that $x/x(\m)$ and $x/x(\fdh)$
in hot cores are relatively lower than in hot corinos, showing that complex 
molecules are relatively more important in hot corinos, a difference that can be explained in the 
case of grain-surface synthesis.

Overall, although there is no absolute proof, there is 
circumstantial evidence for the formation of
complex organic molecules on grain surfaces, either in the cold phase preceding the begin of the 
star formation process, or in the warm-up phase following the birth of the protostar.
In this scenario, there is also a possibility that X-rays emitted by low-mass protostars 
participate in the formation of these complex molecules.
Finally, it clearly stands out from this work that not only more data (single-dish, interferometric, and high-energy transitions) are needed, but also that laboratory studies of grain-surface reactions are necessary 
to answer of the question of which, why, where, and how complex molecules are formed.

\bigskip

{\it Acknowledgments.}
We wish to thank Miguel Requena-Torres and his co-authors, as well as Robin Garrod and Eric
Herbst for communicating the results of
their work prior to publication (at the time of submission).
We are also very grateful to Malcom Walmsley and the anonymous referee for their
careful reading of the manuscript and for suggestions that helped to improve the paper.

\bigskip

\bibliographystyle{aa}
\bibliography{/Users/sandrine/Documents/biblio/apj-jour,/Users/sandrine/Documents/biblio/bib_sb}

%%%%%%%%%%%%%%%%%%%%%%%%%%%%%%%%%%
%%%%%%%%%%%%%%%%%%%%%%%%%%%%%%%%%%

\begin{appendix}

\section{Theories of complex molecules formation \label{formation}}

In this appendix, we review the different reactions that have been
proposed for the formation of complex species. These reactions can
occur in the gas-phase or on grain surfaces.  Gas-phase reactions
usually involve \fdh\ or \m\ as precursors, or parents. These are
known to form on grain surfaces
\citep[e.g.,][]{tielens+hagen82,charnley-etal97} and evaporate in the
gas-phase where they undergo further reaction to form more complex, or
daughter, molecules.  Grain-surface reactions generally consist of H
or O additions and radical-radical reactions.

\subsection{Formic acid\label{hcooh-formation}}

Both gas-phase and grain-surface processes have been proposed for the
formation of HCOOH. In the gas-phase model of \citet{leung-etal84},
the precursor ion of HCOOH, HCOOH$_2^+$, is produced via the radiative
association
\begin{equation}
{\rm H_2O + HCO^+ \rightarrow HCOOH_2^+ + h\nu}
\end{equation}
followed by dissociative electron recombination to produce HCOOH. 
However, \citet{irvine-etal90} indicate that they believe that this reaction does not occur due to a competitive exothermic reaction to form 
H$_3$O$^+$ and CO.
Instead, they attribute the gas-phase formation of HCOOH 
to occur via an
ion-molecule reaction followed again by a dissociative electron
recombination
\begin{eqnarray}
&&{\rm CH_4 + O_2^+ \rightarrow HCOOH_2^+ + H} \label{hcooh-path1}\\
&&{\rm HCOOH_2^+ + e^- \rightarrow HCOOH + H.} \label{hcooh-path2}
\end{eqnarray}
Reaction (\ref{hcooh-path1}) has been measured in the laboratory and
found to be quite rapid at low temperature
\citep{rowe-etal84}. Moreover, the calculated HCOOH abundance agrees
with the values observed by \citet{irvine-etal90} in the dark cloud
L134 N.\\

On grain surfaces, \citet{tielens+hagen82} proposed the formation of
HCOOH through successive additions of H, O, and H to solid-state CO:
\begin{eqnarray}
&&{\rm CO \harrow HCO \oarrow HCOO \harrow HCOOH}
\end{eqnarray}
However, radiolysis experiments (simulating the processing of
interstellar ices by cosmic rays) by \citet{hudson+moore99} showed
that HCOOH could form in H$_2$O--CO mixed ices via the following
sequence:
\begin{eqnarray}
&&{\rm H + CO \rightarrow HCO}\\
&&{\rm HCO + OH \rightarrow HCOOH.}
\end{eqnarray}
The models of \citet{hasegawa+herbst93} are able to reproduce the
abundances observed in massive hot cores for ages larger than
$10^5-10^6$ yr.  Finally, the grain-surface formation of HCOOH is
supported by the interferometric observations of Sgr B2 and W51 by
\citet{liu-etal01}, and would also be consistent with observations of
this molecule in the ices surrounding the massive protostar W33A
\citep{schutte-etal99, gibb-etal00-icesw33}.

\subsection{Methyl formate \label{mf formation}}

The commonly accepted formation path for HCOOCH$_3$ starts with the
reaction between protonated methanol and formaldehyde to form
protonated methyl formate and molecular hydrogen:
\begin{equation}\label{path1}
{\rm [CH_3OH_2]^+ + H_2CO \rightarrow [HC(OH)OCH_3]^+ + H_2}
\end{equation}
followed by dissociative recombination of [HC(OH)OCH$_3$]$^+$ with
electrons to form HCOOCH$_3$ \citep{blake-etal87}.  However,
laboratory and theoretical work by \citet{horn-etal04} indicates the
existence of a very large activation energy for reaction
(\ref{path1}), so that the later cannot lead to the formation of
protonated methyl formate.  Therefore, the formation of methyl formate
in hot cores cannot occur via this reaction.  \citet{horn-etal04}
searched for more favorable transitions between the reactants and
products of reaction (\ref{path1}), but were unsuccessful.  These
authors also investigated reactions involving other abundant species
in hot cores, such as protonated formaldehyde and CO.  They show that
none of the studied processes produces enough methyl formate to
explain the observed abundances.  However, they also state that one
possibility for producing more methyl formate is that formic acid
would be synthesized on grain surfaces and desorbed into the gas
phase, in which case the reaction:
\begin{equation}\label{path2}
{\rm CH_3^+ + HCOOH \rightarrow HC(OH)OCH_3^+~+}~h\nu
\end{equation}
would play a significant role. \\
The downfall is that \citet{horn-etal04} find that even if HCOOH were
injected with an abundance one order of magnitude higher than observed
in OMC-1, their model still predicts a \mf\ abundance between one and
two orders of magnitude below the observed value in this source.
Overall, in the light of their work, \citet{horn-etal04} conclude that no
gas-phase route seem able to reproduce the observed abundances of \mf,
and hence that this molecule should be produced, at least in part, on
grain surfaces.
As pointed out in Sect. \ref{hcooh-formation}, there is additional
evidence in favor of HCOOH being synthesized on grain surfaces.  \\

Two schemes for grain-surface formation of \mf\ have been proposed,
but none of them has undergone laboratory investigation yet:
\begin{itemize}
\item Formation from precursors CO, O, C, and H landing on grain \citep{herbst05}:
\begin{eqnarray}
&&{\rm CO + H \rightarrow HCO}\\
&&{\rm C  \harrow CH \harrow CH_2 \harrow CH_3 \oarrow CH_3O}\\
&&{\rm CH_3O + HCO \rightarrow HCOOCH_3.}\label{rad-rad-hcooch3}
\end{eqnarray}
\citet{charnley+rodgers05} mention that many radicals (like CH$_3$O
and HCO) could form in close proximity via the hot secondary electron
generated by the passage of a cosmic ray through the ice.  In this
case, it would ensue that radical-radical reactions such as
(\ref{rad-rad-hcooch3}) could occur efficiently.
\item \citet{sorrell01} proposed a model in which the photoprocessing
of grain mantles by UV starlight creates a high concentration of
radicals in the bulk interior of mantles. Grain-grain collisions then
provide excess heat causing radical-radical reactions to occur and
form large organic molecules. In this scheme, \mf\ would be produced
from the reaction between the carboxyl acid (COOH) and the methyl
group (CH$_3$) in the following way:
\begin{eqnarray}
&& {\rm CO + OH \rightarrow COOH~~or~~HCO + HCO \rightarrow COOH + CH}\\
&& {\rm CH \harrow CH_2 \harrow CH_3}\\
&& {\rm COOH + CH_3 \rightarrow HCOOCH_3.}\label{rad-rad-hcooch3-2}
\end{eqnarray}
However, any grain-surface chemistry preceding the hot core/corino
phase would occur in a very dense and highly visually extinct
environment, hence well shielded from UV starlight. Therefore, such a
UV photolysis of grains is unlikely to happen, as pointed out by
\citet{peeters-etal06}. Nevertheless, the radical-radical reaction
(\ref{rad-rad-hcooch3-2}) could still be a possible formation path for
\mf\ via cosmic ray processing.
\end{itemize}

\subsection{Dimethyl ether\label{ch3och3-formation}}

\dime\ was proposed by \citet{blake-etal87} to form in the gas-phase by methyl cation transfer to methanol, followed by electron dissociative recombination:
\begin{eqnarray}
&&{\rm CH_3OH_2^+~+~CH_3OH~\rightarrow~CH_3OCH_4^+~+~H_2O}
\label{ch3och3-part1} \\ &&{\rm CH_3OCH_4^+~+~e^-
\rightarrow~CH_3OCH_3~+~H.} \label{ch3och3-part2}
\end{eqnarray}
\citet{peeters-etal06} claim that their models support such a
gas-phase route if the methanol abundance is of the order 10$^{-6}$ or
more, as it is the case for the OMC-1 hot core (recall that for hot
corinos, $X$(\m)~$\le 3 \times 10^{-7}$). However,
\citet{ceccarelli-etal06} note that experimental measurements of the
neutral products of dissociative recombination reactions show that
two-body products such as in reaction (\ref{ch3och3-part2}) are often
minor channels.  Therefore, gas-phase formation of \dime\ is plausible
but not demonstrated.\\

On the grains, \dime\ could be produced by a similar scheme as for
\mf, that is cosmic ray processing followed by the radical-radical
reaction CH$_3$ + CH$_3$O $\rightarrow$ \dime\ \citep{allen+robinson77}. As in
the case of \mf, this reaction has not been validated by laboratory
studies.\\

\subsection{Methyl and ethyl cyanide \label{n-bearing}}

For completeness, we mention here the possible formation routes of
these two molecules.  However, we will not discuss them any further in
this study, due to the lack of data on potential parents such as
\ammonia.
Two substantially different formation routes for CH$_3$CN have been
proposed in the literature: either in the gas-phase or on grain
surfaces.

\begin{itemize}

\item In the \citet{rodgers+charnley01} chemical model of massive hot
cores, \mc\ is synthesized from \ammonia\ in the following way: HCN is
synthesized from the reaction between \ammonia\ and C$^+$ (yielding
HCNH$^+$, reaction (\ref{nh3+c})), followed by electron recombination
or proton transfer to ammonia (\ref{hcnh+}). \mc\ is then formed from
the radiative association between the methyl ion and HCN
(\ref{ch3+hcn}), again followed by electron recombination
(\ref{ch3cn}):
\begin{eqnarray}
&&{\rm NH_3 + C^+ \rightarrow HCNH^+} \label{nh3+c}\\
&&{\rm HCNH^+ + e^- \rightarrow HCN + H~or~HCNH^+ \rightarrow HCN + NH_4^+} \label{hcnh+}\\
&&{\rm CH_3^+ + HCN \rightarrow CH_3CNH^+ + h\nu} \label{ch3+hcn}\\
&&{\rm CH_3CNH^+ + e^- \rightarrow CH_3CN + H.} \label{ch3cn}
\end{eqnarray}
Comparisons between the results from this model and observations suggest that a grain-surface
formation of \mc\ is not required.\\

\item On grain surfaces, \mc\ can be formed by successive hydrogenation of C$_2$N or
by recombination between CN and CH$_3$:
  \begin{itemize}
  \item[$\bullet$]
  \begin{math}
  {\rm C_2N \harrow HCCN \harrow CH_2CN \harrow CH_3CN}.
  \end{math}
 This set of reactions, used in \citet{caselli-etal93}, underestimates
 the \mc\ abundance by a factor about 50 compared to the abundance
 observed in the Orion hot core. This could be explained by the fact
 that HCCN formation is in competition with C$_3$N formation, and the
 former is indicated by the authors as a less important pathway.

  \item[$\bullet$]
  \begin{math}
  {\rm CN + CH_3 \rightarrow CH_3CN}
  \end{math}
  \citep{hasegawa+herbst93}, which yields a \mc\ abundance in good agreement with the Orion hot core
  value for an age of 10$^5$ yr.
  \end{itemize}

\end{itemize}

Regarding C$_2$H$_5$CN, some studies point towards grain-surface
formation of this molecule by hydrogenation of HC$_3$N
\citep{blake-etal87,charnley-etal92,caselli-etal93}.  Observations by
\citet{liu+snyder99} are consistent with this theory.  However, the
abundance ratio of \mc\ to \ec\ predicted by \citet{caselli-etal93} is
at least two orders of magnitude smaller than the ratio observed in
hot cores and hot corinos. Therefore, it seems unlikely that both \mc\
and \ec\ form via the surface reactions proposed by
\citet{caselli-etal93}.\\

\subsection{Timescales for gas-phase processes \label{timescales}}

For the molecules for which no strong argument has been proposed
against gas-phase formation, e.g., \dime, we can estimate a
formation timescale from the slowest rate coefficient of the reactions
involved.  In general, for a reaction between reactants $r_1$ and
$r_2$ yielding a product $p$, we have:
\begin{equation}
\frac{dn_p}{dt}~=~k_f~n_{r_1}~n_{r_2},
\end{equation}
where $k_f$ is the reaction rate coefficient in \ccs.  
In the event that
$n_{r_1}~>>~n_{r_2}$ (which is the case for \dime, as we will see
further down), 
we also have $n_{r_1}~+ ~n_p~\sim~n_{\rm tot}$,
where $n_{\rm tot}$ is defined as $n_{r_1}~+ ~n_{r_2}~+ ~n_p$.
Then,
\begin{eqnarray}
\frac{dn_p}{dt} & = & k_f~(n_{\rm tot}~-~n_p)~n_{r_2} \\
                           & = & -k_f~n_{r_2}~n_p~+~k_f~n_{\rm tot}~n_{r_2} \\
                           & = & -\frac{1}{\tau_f}~n_p~+~k_f~n_{\rm tot}~n_{r_2},
\end{eqnarray}
which has solution of the form $n_p(t)~=~C_1~\exp[-t/\tau_f]~+~C_2$,
where $C_1$ and $C_2$ are constants. The formation timescale is
\begin{eqnarray}
\tau_f & = & (k_f~n_{r_2})^{-1} \\
           & = & (k_f~n_{{\rm H}_2}~x_{r_2})^{-1} \\
           & = & \left(\frac{k_f}{10^{-10}}\right)^{-1}~\left(\frac{n_{{\rm H}_2}}{10^8}\right)^{-1}~
           \left(\frac{x_{r_2}}{10^{-10}}\right)^{-1}~3\times10^4~{\rm yr,} \label{eq timescale}
\end{eqnarray}
where $n_{{\rm H}_2}$ is the H$_2$ density in \cc\ and $x_{r_2}$ is
the abundance of $r_2$, the least abundant reactant.

For \dime, the slowest reaction is (\ref{ch3och3-part1}), which has a
reaction rate of $1\times10^{-10}$~\ccs\ (from the gas-phase chemical
network
osu.2005\footnote{http://www.physics.ohio-state.edu/$\sim$eric/research.html}),
and (predicted) $x_{r_2}\,=\,x_{{\rm CH}_3{\rm
OH}_2^+}\,\sim\,(1-5)\times10^{-10}$
\citep[from][]{rodgers+charnley03}, which is much smaller than
$x_{r_1}\,=\,x_{\rm CH_3OCH_3}\,\gtsim\,1\times10^{-8}$ (from Table \ref{abundances}),
hence justifying the derivation and use of Eq. \ref{eq timescale}. 
For our sources, $\hdensity\,\sim\,10^8-10^9$~\cc, which yields
$\tau_f~=~6\times10^2-3\times10^4$~yr.\\ 
Regarding the destruction of the studied species in the gas-phase, we
expect ions such as H$_3$O$^+$, HCO$^+$, and H$_3^+$ to be the main
destroyers of complex organic molecules during the hot corino phase,
with H$_3$O$^+$ being the most important of the three due to the
evaporation of water contained in the icy grain mantles
\citep{ceccarelli-etal96,rodgers+charnley03}. In this case, we have:
\begin{eqnarray}
\frac{dn_p}{dt} & = & -k_d~n_p~n_{{\rm H}_3{\rm O}^+} \\
                           & = & -\frac{1}{\tau_d}~n_p \\
\Rightarrow n_p(t) & = & C_3~\exp(-t/\tau_d),
\end{eqnarray}
where $k_d$ is the destruction rate, $\tau_d$ is the destruction timescale, and $C_3$ is a constant.
Then,
\begin{eqnarray}
\tau_d & = & (k_d~n_{{\rm H}_3{\rm O}^+})^{-1} \\
           & = & (k_d~n_{{\rm H}_2}~x_{{\rm H}_3{\rm O}^+})^{-1} \\
           & = & \left(\frac{k_d}{10^{-10}}\right)^{-1}~\left(\frac{n_{{\rm H}_2}}{10^8}\right)^{-1}~
           \left(\frac{x_{{\rm H}_3{\rm O}^+}}{10^{-10}}\right)^{-1}~3\times10^4~{\rm yr.}
\end{eqnarray}
Destruction of \dime\ by H$_3$O$^+$ has $k_d~=~10^{-9}$~\ccs\
(osu.2005) and $x_{{\rm H}_3{\rm
O}^+}~=~7\times10^{-11}-5\times10^{-9}$ (from
\citealt{ceccarelli-etal96} for the lowest value, and
\citealt{rodgers+charnley03} and \citealt{stauber-etal05} for the highest
value), which yields $\tau_d~=~6-4\times10^3$~yr. This is comparable
to the mid-range of the formation scale. Moreover, as the protostar
ages, its luminosity increases, causing the expansion of the
evaporation front and the replenishment of the complex molecules,
whether they are mantle species or whether they form from freshly
evaporated \m\ and/or \fdh. Therefore, the destruction by H$_3$O$^+$
is likely compensated by the evaporation or formation of more complex
molecules.\\

\end{appendix}

\end{document}